\documentclass[letterpaper, submit]{AAS}			
\usepackage{bm}
\usepackage{amsmath}
\usepackage{amssymb}
\usepackage[colorlinks=true, pdfstartview=FitV, linkcolor=black, citecolor= black, urlcolor= black]{hyperref}
\usepackage{comment}
\usepackage{empheq}
\usepackage{overcite}
\usepackage{float}
\usepackage{footnpag}			      	

\usepackage{algorithm}
\usepackage{algpseudocode}
\usepackage{fancyhdr}
\usepackage{graphicx}
\usepackage{subcaption}
\usepackage{tabularx}
\usepackage{booktabs}
\usepackage{dirtytalk}

\PaperNumber{25-428}

\begin{document}

\title{Hybrid Particle Gaussian Mixture (H-PGM) Solution for Cislunar Target Tracking}

\author{Ishan P. Paranjape\thanks{Graduate Assistant - Research, Department of Aerospace Engineering, Texas A\&M University, 710 Ross St., College Station, TX 77843.}, Tarun Hejmadi\thanks{Graduate Assistant - Research, Department of Aerospace Engineering, Texas A\&M University, 710 Ross St., College Station, TX 77843.}, Utkarsh Ranjan Mishra\thanks{Sensor Fusion Engineer, Lucid Group, Inc., 7373 Gateway Blvd., Newark, CA 94560} and 
Suman Chakravorty\thanks{Professor, Department of Aerospace Engineering, Texas A\&M University, 710 Ross St., College Station, TX 77843.},
}

\maketitle{} 	


\begin{abstract}

Gauss's method of orbit determination (OD) is one of the most popular, minimal assumption target tracking techniques in astrodynamics, especially for generating an initial state estimate. However, due to Gauss's method's assumption of Keplerian motion (part of the larger two-body problem), this method cannot be applied in a cislunar environment, where three body, non-planar effects dominate. In this work, we showcase a hybrid Particle Gaussian Mixture (H-PGM) filtering method, a purely recursive probabilistic OD framework that relies upon a sequential combination of the Markov Chain Monte Carlo (MCMC) based Particle Gaussian Mixture-II (PGM-II) and Kalman update based Particle Gaussian Mixture-I (PGM-I) filters. This method allows us to fuse probabilistic information with angles-only observations from terrestrial telescopes for short- and long-term cislunar target tracking. This method also allows us to fuse other target \textit{a priori} information in an effort to reduce target uncertainty in the short term. This hybrid filtering technique is demonstrated for several popular and important cislunar orbit regimes and compared with several homogeneous and hybrid filtering frameworks.

\end{abstract}

\section{Introduction}\label{sec:1intro}

The National Cislunar Science and Technology Strategy identifies the extension of space situational awareness (SSA) into a cislunar environment as one of its key objectives. Similar to regimes closer to the Earth's surface, this document describes the need to detect and identify possible colliding objects, space weather elements, and other dangerous spacecraft operations (i.e. tumbling, abrupt maneuvers) to ensure safe spacecraft navigation.\cite{ussf2020} The cislunar domain is defined by the full $4\pi$ steradian volume in which Earth- and Moon-based three-body dynamics dominate.\cite{bolden2022, griggs2023} This volume is almost a thousand times larger than the volume of possible orbits below the geosynchronous (GEO) limit, at which point two-body dynamics become prominent. Many targets operate within this cislunar domain; we refer to them more specifically as cislunar Resident Space Objects (RSOs). In order to maintain a proper catalog of these cislunar RSOs, effective target tracking becomes a critical problem. Interest within the cislunar domain has burgeoned due to missions such as India's Chandrayaan, China's Chang'e-5, and NASA's Artemis. To ensure safe operation of similar targets in a cislunar environment, we use initial orbit determination (IOD) and orbit determination (OD) to initialize a state estimate and continue to maintain custody of such targets.

Every major target tracking problem associated with near-Earth or cislunar regimes -- especially in a probabilistic sense -- begins with IOD, defined as the process of generating an initial state estimate. The goal of IOD for single target tracking is to accumulate several measurements of a single RSO and then to rely on some \textit{a priori} information to fit a state vector through the measurements.\cite{gooding1997} To this day, Gauss's method of orbit determination (OD) -- which consists of IOD and a nonlinear least-squares refinement of the resulting initial state estimate -- continues to be one a powerful method for generating a deterministic initial state estimate, leveraging three consecutive, closely-spaced angular observations.\cite{taff1979} For probabilistic initial orbit determination (IOD) approaches, specialized methods such as probabilistic admissible region (PAR) and constrained admissible region (CAR) persist throughout space situational awareness (SSA) literature.\cite{taff1979, deMars2013, kelecy2013, hussein2014, hussein2014.2, mishra2024} However, most of these probabilistic techniques are limited to the two-body problem and Keplerian mechanics. 

Literature on probabilistic initial orbit determination in a cislunar environment -- encompassing three-body dynamics -- is sparse. Most cislunar orbit determination research involves formulating an initial state estimate whose mean is centered at or close to the true target state (i.e. the truth) and whose state covariance is small.\cite{gupta2025, lopez2025} One set of scientists utilizes sparse grid collocation for generating an initial state estimate by solving a large-scale nonlinear programming (NLP) problem.\cite{heidrich2025} Developing a target tracking technique parallel to the geometric PAR (G-PAR) solution for Keplerian orbits, a team of scientists and engineers extended the idea of PAR to the cislunar domain, postulating information about camera angular rate constraints, light intensity curves for range inference, and Jacobi's constant.\cite{mishra2024, bolden2022, griggs2023} Seeking a way to solve the problem of cislunar initial orbit determination by leveraging as little \textit{a priori} information as possible, we introduced a probabilistic IOD and OD framework leveraging minimal range information or weak range measurements to form sets of initial position estimates, from which initial velocity estimates were derived to form a particle-based initial state estimate.\cite{paranjape2025, paranjape2026} A major conclusion resulting from this work is that even if an IOD method results in a large, highly uncertain initial state estimate, a good OD or filtering technique will properly address this concern.

The second part of target tracking problems involves orbit determination. For probabilistic orbit determination, filters are utilized for uncertainty reduction. The most well-known filter is the Kalman-Bucy Filter, which is typically used for systems with linear dynamics and linear observation models with respective additive, zero-mean Gaussian noises.\cite{kalmanBucy1961} This linearity in dynamics and measurement models is relaxed in the formulations of Extended Kalman Filters (EKFs), Unscented Kalman Filters (UKFs), and Ensemble Kalman Filters (EnKFs).\cite{crassidis2011, wan2000, julier2004, Evensen1994, evensen1995, evensen2003} Within the astrodynamics community, the batch least squares filter -- paired with IOD -- is a popular probabilistic target tracking framework.\cite{tapley2004, mishra2023} For each technique, however, the underlying assumption of Gaussian \textit{a priori} state estimates persists. Due to the nonlinearity of dynamics within the cislunar domain, most Gaussian-based filtering techniques will fail in the absence of strong initial state estimates. Since most \textit{a priori} state estimates may be accurately represented by Gaussian Mixture Models (GMMs), the use of GMMs has grown for cislunar and near-Earth orbit determination problems. Examples of such filters include adaptive GMM filters, ensemble GMM filters, and UKF/particle filter hybrids.\cite{Iannamorelli2025AdaptiveGM, yun2022, popov2023, durant2023, raihanukfpf2018} The UKF/particle filter hybrid formulation, in particular, inspired the creation of the Particle Gaussian Mixture (PGM) filters.\cite{raihan2018, raihan2018pgm2}

The Particle Gaussian Mixture filters refer to two related hybrid filters utilizing particles for estimate prediction, clustering-based interactive mixture modeling for estimating \textit{a priori} state distributions, and a Bayesian update process. At the end of these three steps, a new posterior state estimate is obtained by resampling particles from the updated state estimate. While the propagation and clustering steps are identical for both PGM filters, the two filters -- differentiated as PGM-I and PGM-II -- significantly deviate in their update process. The better-known PGM-I filter was designed to address two major issues: 1) adaptability in the number of GMM components between iterations, and 2) particle depletion resulting from the update step of a particle filter. This filter has been successfully demonstrated for multiple one-dimensional and three-dimensional systems.\cite{raihan2018} The OD framework for G-PAR and its multi-target tracking extension also consisted of this filter.\cite{mishra2023, mishra2024} The Particle Gaussian Mixture filter has also been used to address the problem of cislunar target tracking. The PGM-I filter has proven to be particularly resilient for long sensor shutoffs or long periods without observations, a common occurrence in cislunar target tracking.\cite{bolden2022, griggs2023, paranjape2025, givens2025, paranjape2026}

The PGM-II filter was developed in parallel to address target tracking scenarios in which Kalman updates are insufficient or where a Bayesian update or information fusion expects multi-modality.\cite{raihan2018fusion, raihan2018pgm2} While the PGM-I filter assumes that the number of GMM components of the \textit{a priori} and posterior state estimate are identical, the PGM-II filter adds flexibility in the number of GMM components throughout its update process. To address potential multi-modality in the update process, the PGM-II filter utilizes Markov Chain Monte Carlo (MCMC) sampling. This MCMC sampling step considers an \textit{a priori} GMM estimate and the measurement likelihood function, and samples from the product of these two PDFs, also referred to as the posterior. From here, a second clustering step (which usually changes the number of GMM components) and weight update step complete the update process. This filter has been demonstrated on a limited, but chaotic, set of dynamic systems.\cite{raihan2018pgm2}

The precursor to the orbit determination framework in this article is a combined cislunar IOD/OD framework,  hereafter named the Kinematic Fitting Particle Gaussian Mixture (KF-PGM) framework.\cite{paranjape2025, paranjape2026} The IOD framework in KF-PGM relies upon simply fitting polynomials through a series of observations of the same cislunar RSO along with some weak but valid assumption of the range to estimate target position. By kinematically fitting polynomials or splines through the initial position estimate, an initial state estimate may be formulated and later abstracted into a probabilistic initial state estimate. When this probabilistic initial state estimate is combined with the PGM-I filter subsequently, KF-PGM robustly tracks cislunar RSOs for weeks or even months, even in the event of long sensor shutoffs or gaps between observations.

A disadvantage to the KF-PGM framework, however, is the fact that a large number of observations are required to obtain an initial state estimate. In reality, we may only be able to observe a cislunar RSO for a few times in a week or even a month, causing long lags in the target tracking process, requiring frequent redos of the IOD process. Furthermore, as we shall demonstrate later in this article, although small-order kinematic fits yield a consistent initial state estimate, they break down with the PGM-I filter at subsequent update steps. As a result, a combined IOD/OD framework which leverages minimal assumptions and minimal observations -- similar to Gauss's method -- is desired.

In this article, we present a purely recursive approach that involves both Particle Gaussian Mixture filters for tracking cislunar RSOs. The only two pieces of \textit{a priori} information that this approach requires is that 1) the object belongs to the cislunar domain, and 2) the cislunar RSO is moving at a vaguely valid speed. For the PGM-based update processes, we utilize angles-only measurements from ground-based telescopes. This recursive OD method utilizes a mix of PGM-I and PGM-II based filtering over the same measurement sets as KF-PGM to create up to a 100-fold improvement in state component estimates. Due to the hybrid use of both PGM filters, we refer to this approach as the hybrid PGM (H-PGM) solution. With this approach, we are able to demonstrate an accurate, data-efficient solution to the problem of cislunar target tracking. 

The following sections of this article are organized as follows. First, we set up the cislunar target tracking problem by discussing the dynamics and observation models pertaining to terrestrial-based target tracking. Next, we provide a mathematical background about the PGM-I and PGM-II filters, and Bayesian filtering in general. Then, we discuss a modification in the PGM-II update step that allows us to localize extremely uncertain estimates and fuse other target characteristic information. This hybrid PGM filtering method utilizes both filters sequentially. Finally, we showcase this hybrid PGM filtering approach for popular cislunar orbit regimes, compare various filtering frameworks to our H-PGM solution, and discuss short- and long-term effects of target \textit{a priori} information fusion at the first time step and its effect on subsequent state estimates.

\section{Cislunar Target Tracking Preliminaries}\label{sec:2prelims}

In this article, we define cislunar target tracking as a two-part process. The first part, IOD, is used to generate or abstract a probabilistic initial state estimate for some cislunar RSO. The second part, OD, involves the use of a filter to continuously maintain custody of the same cislunar RSO. We dedicate this section to outlining the dynamics and measurement models which we utilize in this work that lend themselves to making the development of our target tracking framework possible. 

\subsection{Dynamics Model: Circular Restricted Three Body Problem}\label{subsec:cr3bp}

In cislunar regimes, two-body dynamics cannot be assumed because the mass of the Moon becomes too large to ignore. The resulting new gravitational forces lead to non-conic, out of plane motion in an Earth-centered inertial reference frame, violating fundamental assumptions of Keplerian motion, requiring dynamics modeling associated with the three-body problem.\cite{koon1999} Currently, one of the most popular models for explaining three-body dynamics within the cislunar domain, in which the spacecraft or RSO mass is negligible with respect to the Earth and the Moon, is the circular-restricted three-body problem (CR3BP).\cite{holzinger2021, schaub2003} From this dynamics model, several orbital element based approximations have been introduced.\cite{Peterson2023, gupta2025}

The synodic coordinate reference frame in which CR3BP dynamics are expressed is centered at the barycenter (or center of mass) of the two largest orbiting bodies (i.e. the Earth and the Moon). Within the cislunar domain, the Earth and the Moon rotate around their respective barycenter in a circular motion with a period of roughly 29 days. For the Earth-Moon-RSO system, this barycenter is located inside the Earth's surface, and a single period of the Moon's motion around the Earth and barycenter is approximately equal to the length of the lunar cycle. The RSO has negligible mass relative to the Earth and Moon, and its dynamics in the barycentric frame are expressed with respect to the rotation of the Earth-Moon vector. 

\begin{subequations} \label{eq1:cr3bp}
    \begin{align}
        \ddot{x} = x + 2\dot{y} - \frac{(1-\mu)(x+\mu)}{r_1^3} - \frac{\mu x - \mu(1-\mu)}{r_2^3}\label{eq1a:cr3bp_X} \\
        \ddot{y} = y - 2 \dot{x} - \frac{(1-\mu)y}{r_1^3} - \frac{\mu y}{r_2^3}\label{eq1b:cr3bp_Y} \\
        \ddot{z} = \frac{(1-\mu)z}{r_1^3} - \frac{\mu z}{r_2^3} \label{eq1c:cr3bp_Z}
    \end{align}
\end{subequations}
In Eq. \eqref{eq1:cr3bp}, $x$ and $y$ define the Earth-Moon orbital plane, $r_1$ and $r_2$ define the distances between the target and the the Earth and Moon mass centers, respectively. $\mu$ is a constant mass parameter that defines the weight ratio between the middle and most massive bodies. The distance between the Earth and Moon centers is normalized to 1 for simplicity. Time and mass are nondimensionalized accordingly.\cite{schaub2003} 

Eq. \eqref{eq1:cr3bp} consists of five equilibrium points -- three unstable and two stable.\cite{doedel2007, szebehely1969} Interest in the regions around the first two unstable equilibrium points, known as the $L_1$ and $L_2$ Lagrange points, has burgeoned due to their proximity to the Moon and the high potential for chaotic bifurcations and breakdowns in Gaussianity beginning with localized state estimates.\cite{zimovan2017, zimovanspreen2020} In this article, we focus on a few trajectories and orbits around the $L_2$ Lagrange point.

\subsection{Measurement Model}\label{subsec:measModel}

As mentioned in Section \ref{subsec:cr3bp}, the state of the target or cislunar RSO \textit{T} is expressed with respect to the barycenter \textit{B}. However, ground based radars, lidars, and optical telescopes \textit{O} localize observer-target states with angular quantities such as azimuth and elevation in a topocentric reference frame. Letting $\mathbf{x}_{OT}^{\mathcal{T}} = [x^{\mathcal{T}}, y^{\mathcal{T}}, z^{\mathcal{T}}, \dot{x}^{\mathcal{T}}, \dot{y}^{\mathcal{T}}, \dot{z}^{\mathcal{T}}]^T$ define the state vector $\mathbf{x}$ in this topocentric observer frame, we can define our measurement model $\mathbf{z}_k = [AZ, EL]^T$ with the following equations.
\begin{subequations} \label{eq2:AZ-EL}
    \begin{align}
        AZ = tan^{-1} \left(\frac{y^{\mathcal{T}}}{x^{\mathcal{T}}}\right)\label{eq2a:azimuth} \\
        EL = \frac{\pi}{2} - cos^{-1} \left(\frac{z^{\mathcal{T}}}{\sqrt{(x^{\mathcal{T}})^2 + (y^{\mathcal{T}})^2 + (z^{\mathcal{T}})^2}}\right) \label{eq2b:elevation}
    \end{align}
\end{subequations}

For computational efficiency and our article's use of a single observer based in College Station, TX, USA, we elect to express our state vectors exclusively in the topocentric reference frame. The process for converting between the barycentric reference frame to an Earth-centered inertial (ECI) reference frame involves a singular rotation.\cite{dahlke2018} Computational tools may be used to convert between the ECI and topocentric reference frames for a given time. Due to electro-optical (EO) sensors' high fidelity, we assume that the measurement noise for both azimuth and elevation angles is distributed as zero-mean Gaussian with one standard deviation equal to 1.5 arcsec.

Given angles-only measurements, an initial full position estimate requires range information. Obtaining cislunar target ranges using ground-based radars and lidars is difficult, if not impossible, due to the fact that received signal power varies highly inversely with the distance. Although this inaccuracy may be partially modeled by utilizing a high range standard deviation, it is far more efficient and cheaper to utilize weaker forms of range information such as the underlying assumption of an RSO orbiting within the cislunar domain -- which runs a wide gamut of range values.\cite{paranjape2025, paranjape2026} In reality, we may have a good amount of \textit{a priori} information from which we can extrapolate target range and other characteristics. We outline these characteristics in Section \ref{subsec:RSOchars}.

\subsection{Cislunar Target Characteristics}\label{subsec:RSOchars}

We define cislunar target \textit{a priori} information or characteristics to be parameters or traits relating to the state of an object. Some examples of cislunar target \textit{a priori} information includes range, angular rates, potential energy, and Jacobi's constant of integration.\cite{zimovan2017, dahlke2018, zimovanspreen2020} Just like for PAR-based techniques, these orbital characteristics are expressed as a PDF. Since cislunar RSO characteristics are rarely expressed as a measurement with an associated measurement noise but rather as a range with upper and lower bound values, we assume a uniform distribution for target \textit{a priori} information. This also allows us to model maximum possible uncertainty in our \textit{a priori} information. We define $\mathbf{\Theta}_k$ as a multivariate uniform PDF defining sets of target \textit{a priori} information. 

To demonstrate the abstraction of target \textit{a priori} information into a PDF, we use the following example. Within the past decade, research into the 9:2 resonant near-rectilinear Halo orbit (NRHO) in the Earth-Moon system has grown.\cite{2019NASA} Knowledge that a target exists within this orbit regime may be abstracted probabilistically into a set of valid observer-target distances or potential energies over time by a simple simulation. We can place upper and lower bounds on target-observer range and/or orbit energies to abstract a uniform PDF.\cite{zimovan2017, zimovanspreen2020} In theory, the fusion of this PDF with the angles-only measurement set will allow us to better narrow the probabilistically admissible regions of the target states compared to solely the measurements.

We emphasize that cislunar target \textit{a priori} information is \textit{not} considered part of the measurement model in any way. Rather, we fuse these target characteristics sequentially with the measurements. The values of these target characteristics can also be functions of the target state in multiple reference frames. To make that distinction, we denote the state vector in the topocentric and CR3BP synodic reference frames as $\mathbf{x} = [x, y, z, \dot{x}, \dot{y}, \dot{z}]^T$ and $\mathbf{x}^s = [x^s, y^s, z^s, \dot{x}^s, \dot{y}^s, \dot{z}^s]^T$, respectively. This notation is slightly different from those expressed in Sections \ref{subsec:cr3bp} and \ref{subsec:measModel}. We provide the translation models $\mathbf{\theta}(\mathbf{x})$ between target state and target characteristic in both reference frames in Table \ref{table:params}. In the following section, we define the filtering methods that we shall use to fuse target characteristics with our underlying state estimates and measurements.

\begin{table}[h]
\caption{Cislunar Target \textit{A Priori} Information and Functions of State}
\label{table:params}
\begin{center}
\begin{tabular}{|c||c|}
\hline
\textbf{\textit{A Priori} Information} & $\mathbf{\theta(\mathbf{x}})$\\
\hline
Range ($\rho$) & $\sqrt{x^2 + y^2 + z^2}$ \\
\hline
Angular Rates ($[\dot{AZ}, \dot{EL}]^T$) & $[\frac{\dot{y}x-y\dot{x}}{x^2+y^2}, -\frac{\dot{z}(x^2+y^2) - z(x\dot{x}+y\dot{y})}{\rho^2 \sqrt{x^2+y^2}}]^T$\\
\hline
Orbit Energy ($\mathcal{U}^*$) & $\frac{1-\mu}{r_1}+\frac{\mu}{r_2}+\frac{1}{2}[(x^s)^2+(y^s)^2]$\\
\hline
Jacobi's Constant ($JC$) & $2\mathcal{U}^* - [(\dot{x}^s)^2+(\dot{y}^s)^2+(\dot{z}^s)^2]$\\
\hline
\end{tabular}
\end{center}
\end{table}

\section{Particle Gaussian Mixture Filters}\label{sec:pgm}

In this section, we discuss the two Particle Gaussian Mixture filters extensively.\cite{raihan2018, raihan2018pgm2} Although most of the article focuses on the application of the PGM-I and PGM-II filters in a cislunar environment, this section focuses on the general PGM-I and PGM-II filters and is applicable to any system with nonlinear dynamics and nonlinear measurement models.

\subsection{Preliminaries of Bayesian Filtering}

Suppose $\mathbf{x}_k \in \mathbb{R}^{n_x}$ represents an $n_x$-dimensional state vector $\mathbf{x}_k$ at time step $k$. This state vector is propagated according to a nonlinear function $\mathbf{f}(\mathbf{x}_k)$ and a Markov Chain that has a probability density function (PDF) defined by $p(\mathbf{x}_k|\mathbf{x}_{k-1})$ for time steps $k$ and $k-1$. Let $\mathbf{z}_k \in \mathbb{R}^{n_z}$ represent an $n_z$-dimensional vector of observation quantities at time step $k$. $\mathbf{z}_k$ is related to the state vector by a nonlinear function $\mathbf{h}(\mathbf{x})$ with added measurement noise. In most cases, this measurement noise -- expressed as a conditional PDF $p(\mathbf{z}_k|\mathbf{x})$ -- is modeled as additive, zero-mean Gaussian. The general form of the dynamics and measurement equations is given as
\begin{subequations} \label{eq3:dynMeas}
    \begin{align}
        \mathbf{x}_{k+1} = \mathbf{f}(\mathbf{x}_k) + \mathbf{w}_k \label{eq3a:dyn}\\ 
        \mathbf{z}_{k} = \mathbf{h}(\mathbf{x}_k) + \mathbf{v}_k, \label{eq3b:meas}
    \end{align}
\end{subequations}
where $\mathbf{f}(\mathbf{x}_k)$ describes the nonlinear dynamics model, $\mathbf{w}_k$ represents the process noise, and $\mathbf{v}_k$ represents the process noise, which we noted for the PGM Filters is modeled as zero-mean Gaussian. We do not consider measurement noise in this article since the cislunar environment can be relatively free of external disturbances in many cases and because both PGM filter updates tend not to be overconfident.

We can use the transition density function $p(\mathbf{x}_k|\mathbf{x}_{k-1})$ to propagate our PDF with time. Given a sequence of observations $\mathbf{Z}^{k-1}$ up to time $k-1$, we denote the posterior function at time step $k-1$ as $p^{+}(\mathbf{x}_{k-1}) = p(\mathbf{x}_{k-1}|\mathbf{Z}^{k-1})$. The \textit{a priori} distribution for the state at time step $k$ then becomes
\begin{equation}\label{eq4:transitionDensity}
    p^{-}(\mathbf{x}_k) = p^{+}(\mathbf{x}_{k-1})p(\mathbf{x}_k|\mathbf{x}_{k-1}).
\end{equation}

At this point, we slightly abuse notation such that $p^{+}(\mathbf{x}_k) = p_k^{+}(\mathbf{x})$ and $p^{-}(\mathbf{x}_k) = p_k^{-}(\mathbf{x})$. Given an observation $\mathbf{z}_k$ at time step $k$, we apply Bayes' rule such that the posterior state distribution at time step $k$ becomes
\begin{equation} \label{eq5:bayesRule}
    p_{k}^{+}(\mathbf{x}) = \frac{p_k^{-}(\mathbf{x}) p(\mathbf{z}_k|\mathbf{x})}{p(\mathbf{z}_k)} = \frac{p_k^{-}(\mathbf{x}) p(\mathbf{z}_k|\mathbf{x})}{\int p_k^{-}(\mathbf{x}) p(\mathbf{z}_k|\mathbf{x}) d\mathbf{x}}.
\end{equation}
The denominator on the right hand side (RHS) of Eq. \eqref{eq5:bayesRule} is derived from the continuous form of the law of total probability. Both Particle Gaussian Mixture filters assume that $p_k^{-}(\mathbf{x}_k)$ for all $k$ is expressed as a Gaussian Mixture Model (GMM)
\begin{equation} \label{eq6:gmm}
    p_k^{-}(\mathbf{x}) = \sum_{i=1}^{N} \omega_i^{-} \mathcal{N}(\mathbf{x}_k; \mathbf{\mu}_i^{-}, \mathbf{P}_i^{-}),
\end{equation}
with $N$ Gaussian mixture components with weights $\omega_i^{-}$ and normal distributions $\mathcal{N}(\cdot)$ represented by \textit{a priori} means $\mathbf{\mu}_i^{-}$ and \textit{a priori} covariances $\mathbf{P}_i^{-}$. Next, we show that if $p_k^{-}(\mathbf{x})$ is expressed as a GMM, then the posterior estimate $p_k^{+}(\mathbf{x})$ may also be expressed as a GMM.

We substitute Eq. \eqref{eq6:gmm} into the RHS of Eq. \eqref{eq5:bayesRule}, noting that the sum and integral are linear operators.
\begin{equation}\label{eq7:bayesExpansion}
    \begin{aligned}
        p_k^{+}(\mathbf{x}) = \frac{p_k^{-}(\mathbf{x}) p(\mathbf{z}_k|\mathbf{x})}{\int p_k^{-}(\mathbf{x}) p(\mathbf{z}_k|\mathbf{x}) d\mathbf{x}} \\
        = \frac{\sum_{i=1}^{N} \omega_i^{-} \mathcal{N}(\mathbf{x}; \mathbf{\mu}_i^{-}, \mathbf{P}_i^{-}) p(\mathbf{z}_k|\mathbf{x})}{\int \sum_{j=1}^{N} \omega_j^{-} \mathcal{N}(\mathbf{x}; \mathbf{\mu}_j^{-}, \mathbf{P}_j^{-}) p(\mathbf{z}_k|\mathbf{x}) d\mathbf{x}} \\
        = \frac{\sum_{i=1}^{N} \omega_i^{-} \mathcal{N}(\mathbf{x}; \mathbf{\mu}_i^{-}, \mathbf{P}_i^{-}) p(\mathbf{z}_k|\mathbf{x})}{\sum_{j=1}^{N} \omega_j^{-} \int \mathcal{N}(\mathbf{x}; \mathbf{\mu}_j^{-}, \mathbf{P}_j^{-}) p(\mathbf{z}_k|\mathbf{x})d\mathbf{x}}.
    \end{aligned}
\end{equation}
We define the likelihood function $l_i(\mathbf{z}_k)$ -- expressing the probability density value of an observation $\mathbf{z}_k$ coming from the $i$-th \textit{a priori} GMM component -- as 
\begin{equation}\label{eq8:likelihood}
    l_i(\mathbf{z}_k) = \int_{\mathbb{R}^{n_x}} \mathcal{N}(\mathbf{x}; \mathbf{\mu}_i^{-}, \mathbf{P}_i^{-}) p(\mathbf{z}_k|\mathbf{x}) d\mathbf{x}.
\end{equation}
We substitute Eq. \eqref{eq8:likelihood} into Eq. \eqref{eq7:bayesExpansion}, noting that $l_i(\mathbf{z}_k) > 0$ for all $\mathbf{z}_k$.

\begin{equation}\label{eq9:posteriorGMM}
    \begin{aligned}
        p_k^{+}(\mathbf{x}) = \frac{\sum_{i=1}^{N} \omega_i^{-} \mathcal{N}(\mathbf{x}; \mathbf{\mu}_i^{-}, \mathbf{P}_i^{-}) p(\mathbf{z}_k|\mathbf{x})}{\sum_{j=1}^{N} \omega_j^{-} l_i(\mathbf{z}_k)} \frac{l_i(\mathbf{z}_k)}{l_i(\mathbf{z}_k)} \\
        = \sum_{i=1}^{N} \frac{\omega_i^{-} l_i(\mathbf{z}_k)}{\sum_{j=1}^{N} \omega_j^{-} l_i(\mathbf{z}_k)} \frac{\mathcal{N}(\mathbf{x}; \mathbf{\mu}_i^{-}, \mathbf{P}_i^{-}) p(\mathbf{z}_k|\mathbf{x})}{l_i(\mathbf{z}_k)} \\
        = \sum_{i=1}^{N} \omega_i^{+} p_{i,k}^{+}(\mathbf{x}).
    \end{aligned}
\end{equation}
Multiplying the numerator and denominator by $l_i(\mathbf{z}_k)$ and rearranging the terms gives us a product of two components: the discrete (i.e. weight) update and the continuous (i.e. PDF) update. If we assume additive zero-mean Gaussian measurement, then the continuous update $p_{i,k}^{+}(\mathbf{x})$ simply becomes a Kalman update for each component. These discrete and continuous update equations are given by Eqs. \eqref{eq10a:discreteUpdate} and \eqref{eq10b:continuousUpdate}, respectively.
\begin{subequations}\label{eq10:posteriorUpdates}
    \begin{align}
        \omega_i^{+} = \frac{\omega_i^{-}l_i(\mathbf{z_k})}{\sum_{j=1}^{N} \omega_j^{-}l_j(\mathbf{z}_k)} \label{eq10a:discreteUpdate} \\ 
        p_{i,k}^{+}(\mathbf{x}) = \frac{\mathcal{N}(\mathbf{x}; \mathbf{\mu}_i^{-}, \mathbf{P}_i^{-}) p(\mathbf{z}_k|\mathbf{x})}{l_i(\mathbf{z}_k)} \label{eq10b:continuousUpdate} . 
    \end{align}
\end{subequations}

Both PGM filters utilize the same discrete update in Eq. \eqref{eq10a:discreteUpdate}. They differ, however, in terms of likelihood computation. Furthermore, while the PGM-I filter utilizes a Kalman update, the PGM-II filter utilizes an intensive, three-step update process with MCMC sampling, clustering, and weight updates. These differences are explained in greater detail in Sections \ref{subsec:pgm1} and \ref{subsec:pgm2}. 

\subsection{Similarities Between the Particle Gaussian Mixture Filters}\label{subsec:similarities}

Before highlighting major and minor differences between the Particle Gaussian Mixture filters, we outline their similarities. Both Particle Gaussian Mixture filters utilize particles to propagate GMM-based estimates, much like a particle filter. The PGM filters use particle-based propagation due to the ability of particles to better define the breakdown of Gaussianity in the presence of chaotic, nonlinear dynamics. Especially in cases of long measurement separation, particle ensembles may bifurcate and diverge.\cite{zimovan2017, zimovanspreen2020, paranjape2026, givens2025} From a computational standpoint, this is the least efficient part of the PGM-I algorithm. 

The ensemble of particles is propagated until the time of the next observation. These particles are abstracted into a PDF as follows. Since both PGM filters require a GMM-based \textit{a priori} estimate, both PGM filters have to use a clustering algorithm (broadly referred to as $\mathcal{C}$) to fit or approximate a GMM-based estimate. So far, most PGM-based works have utilized the \textit{k-means++} algorithm, while others have utilized DBSCAN.\cite{lloyd1982, kim2026} Our work utilizes the former method. The \textit{k-means++} algorithm partitions the ensemble of propagated particles into $M^-$ different sets or clusters. From the clustered particles, we can compute the weight, mean, and covariance associated with each GMM component. Both PGM filters bypass the problem of particle depletion in particle filters by utilizing a Bayesian update for each GMM component rather than the particles associated with each \textit{a priori} GMM component. After their respective Bayesian updates, the PGM filters resample $N_p$ particles from the posterior GMM-based estimate, which are then propagated per the system dynamics, leading to the next iteration for both filters.

\subsection{Particle Gaussian Mixture-I Filter}\label{subsec:pgm1}

As previously noted, the Particle Gaussian Mixture filters differ in their Bayesian update processes. The first Particle Gaussian Mixture filter -- PGM-I -- focuses on a Kalman or Kalman-family update step, expressed by
\begin{equation}\label{eq11:kalmanUpdate}
    p_{i,k}^{+}(\mathbf{x}) = \frac{\mathcal{N}(\mathbf{x}; \mathbf{\mu}_i^{-}, \mathbf{P}_i^{-}) p(\mathbf{z}_k|\mathbf{x})}{l_i(\mathbf{z}_k)} = \mathcal{N}(\mathbf{x}; \mathbf{\mu}_i^{+}, \mathbf{P}_i^{+}),
\end{equation}
for each GMM component $i$ at time step $k$. Although literature about and adjacent to the PGM-I filter typically involves EKF-based updates for several examples, more recent literature has trended towards an Ensemble Kalman Filter (EnKF) type of update.\cite{raihan2018, raihan2018pgm2, mishra2024, paranjape2025} We describe the PGM-I update process below.

The PGM-I filter starts by considering an initial state estimate that can be accurately modeled as a GMM. The propagation and clustering steps are described in Section \ref{subsec:similarities}. Literature about the PGM-I filter only considers up to three types of update steps for each GMM component -- an EKF update, a UKF update, and an EnKF update. While the former two update steps are well-known, they have a tendency to be overconfident after long gaps between observations. The EnKF update step considered in this article as well as some previous works involving PGM-I filtering is a variation of the original EnKF's update step.\cite{evensen2003, mishra2024} Since the variation of the EnKF considered in this paper is particle-based through the clustering step, the EnKF-based update fuses observations with the \textit{a priori} component distributions (i.e. component means and covariances) rather than the particles of each GMM component. This consideration mitigates overconfidence at the update step. 

\begin{figure}[thpb] 
      \centering
      \begin{subfigure}{\textwidth}
            \centering
            \includegraphics[width=\linewidth]{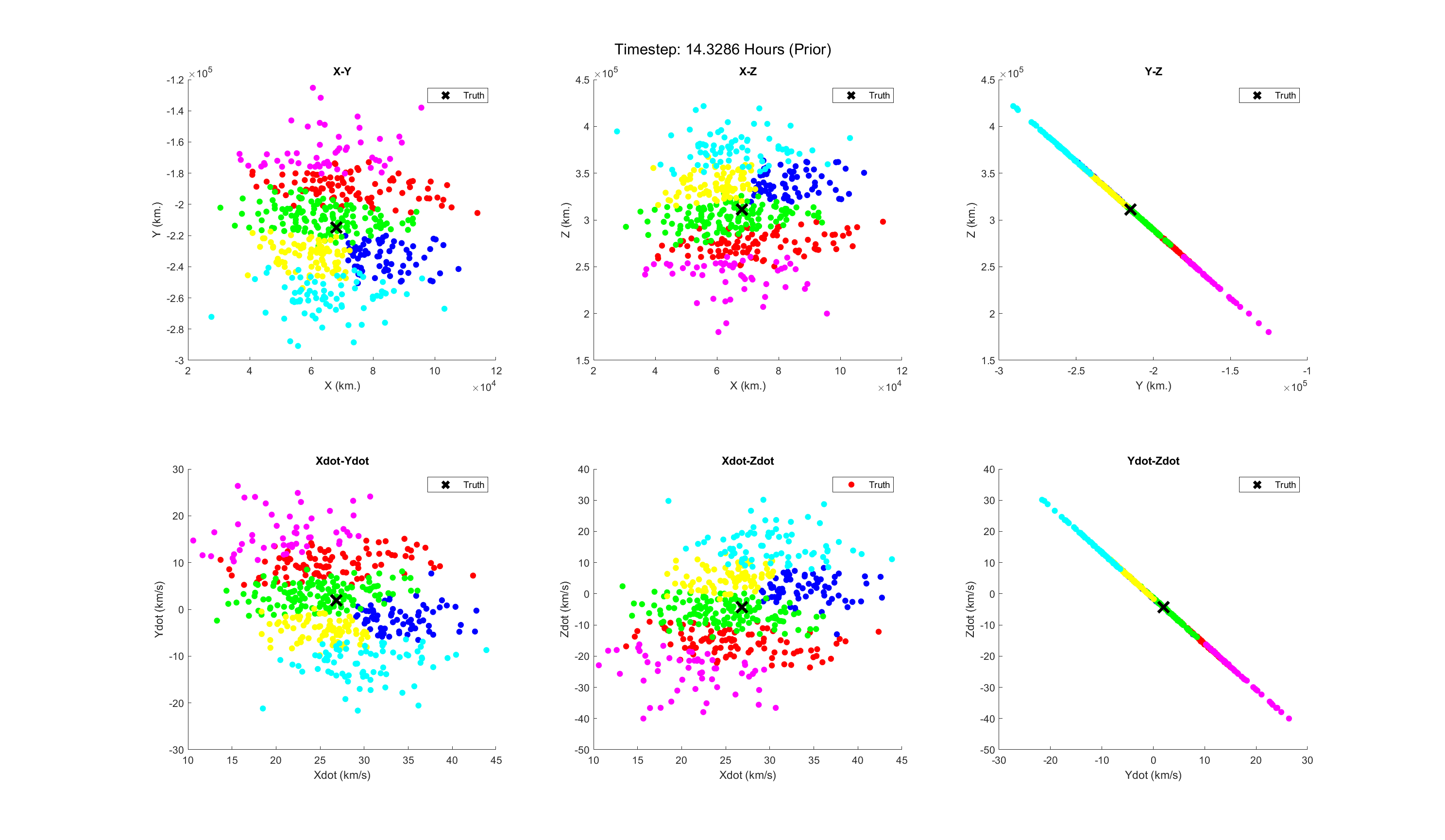}
            \caption{\textit{A priori} estimate for some cislunar RSO clustered into a GMM with six components} \label{fig:1aAprioriEx_Williams}
      \end{subfigure}
      \vspace{0.5cm} 
      \begin{subfigure}{\textwidth}
            \centering
            \includegraphics[width=\linewidth]{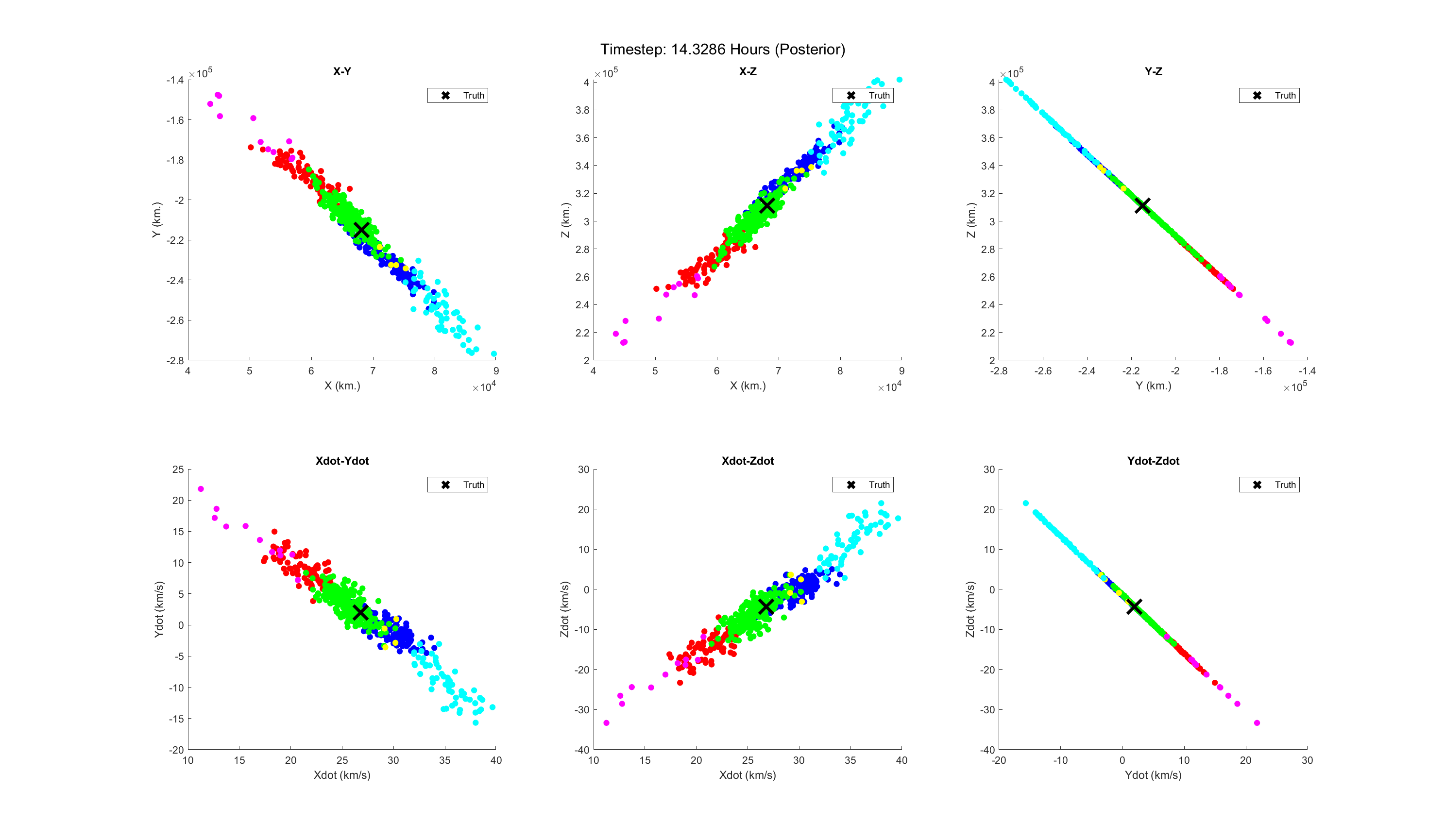}
            \caption{EnKF-based update for each GMM comoponent in Fig. \ref{fig:1aAprioriEx_Williams}}\label{fig:1bPosteriorEx_Williams}
      \end{subfigure}

      \caption{This example is borrowed from the authors' previous work with permission.\cite{paranjape2026} An \textit{a priori} estimate is partitioned into six clusters or GMM components per the PGM-I algorithm. An EnKF-based update is performed directly on each cluster/GMM component. The result is a posterior estimate localized roughly along a straight line, indicative of uncertainty in the target range.}
    \label{fig:1williams2017pgm1Update}
\end{figure}

An illustration of a PGM-I filter's update process, beginning with some \textit{a priori} estimate, is provided by Figure \ref{fig:1williams2017pgm1Update}.\cite{paranjape2026} We also summarize the PGM-I filter in Algorithm \ref{alg:pgm1}.\cite{raihan2018}

\begin{algorithm}
\caption{Particle Gaussian Mixture Filter I}\label{alg:pgm1}
\begin{algorithmic}
    \State Given $\pi_0 (x) = \sum_{i=1}^{M(0)} \omega_i (0) p_g(x; \mu_i (0), P_i (0))$, transition density kernel $p(x'|x)$, $n = 1$.
    \begin{enumerate}
        \State Sample $N_p$ particles from $\pi_{n-1}$ and the transition density kernel $p_n (x'|x)$ as follows: \label{alg:repeat1}
        \begin{itemize}
            \item Sample $X^{(i)^\prime}$ from $\pi_{n-1} (.)$.
            \item Sample $X^{(i)}$ from $p(.|X^{(i)^\prime})$.
        \end{itemize}
        \State Use a clustering algorithm $\mathcal{C}$ to cluster the set of particles $X^{(i)}$ into $M^{-} (n)$ Gaussian clusters with weights, mean, and covariance given by $\{\omega_i^{-} (n), \mu_i^{-} (n), P_i^{-} (n)\}$.
        \State Update the mixture weights and mixture means and covariances to $\{\omega_i^{+} (n), \mu_i^{+} (n), P_i^{+} (n)\}$, given the observation $z_n$, utilizing an EKF, UKF, or EnKF-based Kalman update.
        \State $n = n + 1$. Go to Step \ref{alg:repeat1}.
    \end{enumerate}
\end{algorithmic}\end{algorithm}

\subsection{Particle Gaussian Mixture-II Filter}\label{subsec:pgm2}

The propagation and clustering processes for both the PGM-I and PGM-II filters are identical. The key distinction between the Particle Gaussian Mixture-I and Particle Gaussian Mixture-II filters lies within the update process. In certain cases, a Kalman-based update can be insufficient, and multi-modality at the update step is expected.\cite{raihan2018pgm2} The PGM-II filter was created to address the issue of inflexibility of the number of GMM components between the clustering and update steps of the PGM-I filter. The PGM-II filter's update process consists of three major steps: 1) Markov Chain Monte Carlo (MCMC) sampling, 2) clustering of resulting MCMC chain samples, and 3) weight update. 

The first part of the PGM-II filter update involves MCMC sampling. A sampling function expressed as 
\begin{equation}\label{eq12:MCMCsamplingfunc}
    P(\mathbf{x}) = p_k^{-}(\mathbf{x})p(\mathbf{z}_k|\mathbf{x})
\end{equation}
is used. We denote this posterior sampling function with a capital $P$ to emphasize the fact that it is not a proper PDF, but rather a function proportional to Bayes rule in Eq. \eqref{eq5:bayesRule}. Since the traditional PGM-II filter algorithm defines every \textit{a priori} estimate as a Gaussian Mixture Model, the apparent posterior sampling function would be expressed as 
\begin{equation}\label{eq13:Apparentsamplingfunc}
    P(\mathbf{x}) = \sum_{i=1}^{N} \omega_i^{-} \mathcal{N}(\mathbf{x}; \mathbf{\mu}_i^{-}, \mathbf{P}_i^{-})p(\mathbf{z}_k|\mathbf{x}).
\end{equation}

In this work, however, we do not use Eq. \eqref{eq13:Apparentsamplingfunc} because it causes MCMC sampling to be more computationally intensive than it already is. For this reason, it is recommended that MCMC sampling be done component by component.\cite{raihan2018pgm2} Mathematically, each posterior sampling function would be expressed as 
\begin{equation}\label{eq14:componentSamplingFunc}
    P_i(\mathbf{x}) = \mathcal{N}(\mathbf{x}; \mathbf{\mu}_i^{-}, \mathbf{P}_i^{-})p(\mathbf{z}_k|\mathbf{x}),
\end{equation}
for $i = 1, 2, \cdots M^-$. The \textit{a priori} component weights $\omega_i^{-}$ are retained for the weight update. 

The Metropolis algorithm is commonly used for MCMC sampling since it uses a symmetric proposal PDF, whose spread we control such that the acceptance rate of the Metropolis algorithm is ideally between 10-40\%.\cite{mishra2021} In the original PGM-II work, the authors recommend utilizing a proposal function of the form 
\begin{equation}\label{eq14:componentSamplingFunc}
    q_i(\mathbf{x}) = \mathcal{N}(\mathbf{x}; \mathbf{\mu}_i^{-}, K_p\mathbf{P}_i^{-}),
\end{equation}
where $i$ denotes a specific \textit{a priori} GMM component, and $K_p$ is a user-controlled parameter quantifying proposal spread.\cite{raihan2018pgm2} To efficiently sample the state space, we recommend running hundreds of MCMC chains in parallel. These numerous chains result in a large ensemble of particles representative of the posterior sampling functions for each \textit{a priori} GMM component. This particle ensemble is denoted by $A_i$.

The second step of the PGM-II filter update process involves clustering. The same clustering algorithm $\mathcal{C}$ as the PGM-I filter's clustering process may be used for each component's MCMC chain ensemble. We primarily cluster particles within $A_i$ to de-correlate samples of the MCMC chains due to their random walk behavior. The MCMC chains from the first part of the update process may be further de-correlated by picking every thousandth or so sample from each of the parallel Markov chains beyond a certain burn-in period. Sequentially picking every thousandth or so particle for each MCMC chain in $A_i$ and then clustering using $\mathcal{C}$ results in a PDF which we can denote as
\begin{equation}\label{eq15:clustPDF}
    p_{i,m}(\mathbf{x}) = \sum_{m=1}^{N_i} \omega_{i,m}^{+} \mathcal{N}(\mathbf{x}; \mathbf{\mu}_{i,m}^{+}, \mathbf{P}_{i,m}^{+}).
\end{equation}
The number of clusters $N_i$ can vary with each MCMC-based ensemble $A_i$ from the first part of this update process, depending on the shape of the ensemble. Furthermore, $\omega_{i,m}^{+}$ denotes the cluster weight and this value will need to be multiplied by the posterior ensemble weight. We denote the $(i,m)$ subscript and $+$ superscript for the mean and covariance quantities of Eq. \eqref{eq15:clustPDF} to underscore the actual posterior component PDFs. We only need to compute the ensemble weight update now.

This second clustering step offers much more flexibility than the PGM-I filter during the update process because the number of clusters at this particular step may be user-defined. Since this clustering step is done for each $A_i$, the number of GMM components in the posterior estimate is strictly greater than or equal to the number of GMM components from the \textit{a priori} mixture. If we desire fewer GMM components than the \textit{a priori} estimate at the end of the clustering process, it would make sense to use the apparent posterior sampling function in Eq. \eqref{eq13:Apparentsamplingfunc}. However, since Eq. \eqref{eq13:Apparentsamplingfunc} is computationally unwieldy, each posterior estimate considered in this work has an equal or greater number of GMM components as its \textit{a priori} estimate. We shall next discuss the PGM-II filter's weight update.

The third and final step of the PGM-II update process computes the posterior component weights. We set the \textit{a priori} weight for each $A_i$ to be equal to the \textit{a priori} weight of its corresponding GMM component. Just like the PGM-I filter, we use Eq. \eqref{eq10a:discreteUpdate} to update ensemble weights. However, computation of $l_i(\mathbf{z}_k)$ differs. 

We begin with Bayes' rule in Eq. \eqref{eq10b:continuousUpdate} for each GMM component and by considering any proper PDF $\pi (\mathbf{x}) > 0$.
\begin{equation}\label{eq16:componentBayes}
    p_{i,k}^{+}(\mathbf{x}) = \frac{\mathcal{N}(\mathbf{x}; \mathbf{\mu}_i^{-}, \mathbf{P}_i^{-}) p(\mathbf{z}_k|\mathbf{x})}{l_i(\mathbf{z}_k)}\frac{\pi (\mathbf{x})}{\pi (\mathbf{x})} = \mathcal{N}(\mathbf{x}; \mathbf{\mu}_i^{+}, \mathbf{P}_i^{+})\frac{\pi (\mathbf{x})}{\pi (\mathbf{x})}.
\end{equation}
Rearranging these fractions, we obtain
\begin{equation}\label{eq17:cuExpansion2}
        \frac{l_{i}(\mathbf{z}_k)}{\pi(\mathbf{x})} = \frac{\mathcal{N}(\mathbf{x}; \mathbf{\mu}_i^{-}, \mathbf{P}_i^{-}) p(\mathbf{z}_k|\mathbf{x})}{p_{i,k}^{+}(\mathbf{x}) \pi(\mathbf{x})}.
\end{equation}
Then, we take the inverse of both sides, assuming all of the quantities in Eq. \eqref{eq17:cuExpansion2} are strictly positive for all inputs.
\begin{equation}\label{eq18:cuExpansion3}
        \frac{\pi(\mathbf{x})}{l_{i}(\mathbf{z}_k)} = \frac{p_{i,k}^{+}(\mathbf{x}) \pi(\mathbf{x})}{\mathcal{N}(\mathbf{x}; \mathbf{\mu}_i^{-}, \mathbf{P}_i^{-}) p(\mathbf{z}_k|\mathbf{x})}.
\end{equation}
We integrate both sides with respect to $d\mathbf{x}$, which simplifies the left hand side (LHS) of Eq. \eqref{eq18:cuExpansion3} considerably.
\begin{equation}\label{eq19:cuExpansion4}
    \begin{aligned}
        \int_{\mathbb{R}^{n_x}} \frac{\pi(\mathbf{x})}{l_{i}(\mathbf{z}_k)} d\mathbf{x} = \int_{\mathbb{R}^{n_x}} \frac{p_{i,k}^{+}(\mathbf{x}) \pi(\mathbf{x})}{\mathcal{N}(\mathbf{x}; \mathbf{\mu}_i^{-}, \mathbf{P}_i^{-}) p(\mathbf{z}_k|\mathbf{x})} d\mathbf{x} \\
        = \frac{1}{l_i(\mathbf{z}_k)} = \int_{\mathbb{R}^{n_x}} \frac{p_{i,k}^{+}(\mathbf{x}) \pi(\mathbf{x})}{\mathcal{N}(\mathbf{x}; \mathbf{\mu}_i^{-}, \mathbf{P}_i^{-}) p(\mathbf{z}_k|\mathbf{x})} d\mathbf{x}.
    \end{aligned}
\end{equation}

At this point, we consider the fact that each component ensemble $A_i$ is discrete. We manipulate the continuous-form result in Eq. \eqref{eq19:cuExpansion4} to discretize the likelihood computation by recognizing the following expectation property:
\begin{equation}\label{eq20:discreteLikelihood}
    \begin{aligned}
        \frac{1}{l_i(\mathbf{z}_k)} = \int_{\mathbb{R}^{n_x}} \frac{p_{i,k}^{+}(\mathbf{x}) \pi(\mathbf{x})}{\mathcal{N}(\mathbf{x}; \mathbf{\mu}_i^{-}, \mathbf{P}_i^{-}) p(\mathbf{z}_k|\mathbf{x})} d\mathbf{x} \\
        = \mathbb{E}\left[ \frac{\pi(\mathbf{x})}{p(\mathbf{z}_k|\mathbf{x}) \mathcal{N}(\mathbf{x}; \mathbf{\mu}_i^{-}, \mathbf{P}_i^{-})} \right] \\
        \approx \frac{1}{S_i} \sum_{s=1}^{S_i} \frac{\pi(\mathbf{x}_s)}{p(\mathbf{z}_k|\mathbf{x}_s) \mathcal{N}(\mathbf{x}_s; \mathbf{\mu}_i^{-}, \mathbf{P}_i^{-})}.
    \end{aligned}
\end{equation}

In Eq. \eqref{eq20:discreteLikelihood}, $\mathbf{x}_s$ refers to each sample within $A_i$ (after removing burn-in samples and de-correlating remaining samples). While any proper PDF $\pi(\mathbf{x}_s)$ is theoretically acceptable, we simplify Eq. \eqref{eq20:discreteLikelihood} by equating $\pi(\mathbf{x}_s) = \mathcal{N}(\mathbf{x}_s; \mathbf{\mu}_i^{-}, \mathbf{P}_{i}^{-})$. Once we compute all ensemble likelihoods with Eq. \eqref{eq20:discreteLikelihood}, the resulting posterior ensemble weights $\omega_i^{+}$ are multiplied by re-clustered weights $\omega_{i,m}^{+}$ from Eq. \eqref{eq15:clustPDF} for all $i$ and $(i,m)$ tuple to obtain the posterior weights for the $(i,m)$-th GMM component. The final posterior PDF $p^+(\mathbf{x})$ then becomes 
\begin{equation}\label{eq21:posteriorPDF}
        p^+(\mathbf{x}) = \sum_{i=1}^{N}\sum_{m=1}^{N_i}\omega_i^{+}\omega_{i,m}^{+}\mathcal{N}(\mathbf{x}; \mathbf{\mu}_{i,m}^{+}, \mathbf{P}_{i,m}^{+}).
\end{equation}
We draw numerous particles from Eq. \eqref{eq21:posteriorPDF}, leading to the next PGM-II filtering iteration.

\begin{figure}[!thpb]
    \centering
    \begin{subfigure}{\textwidth}
        \centering
        \includegraphics[width=0.5\linewidth]{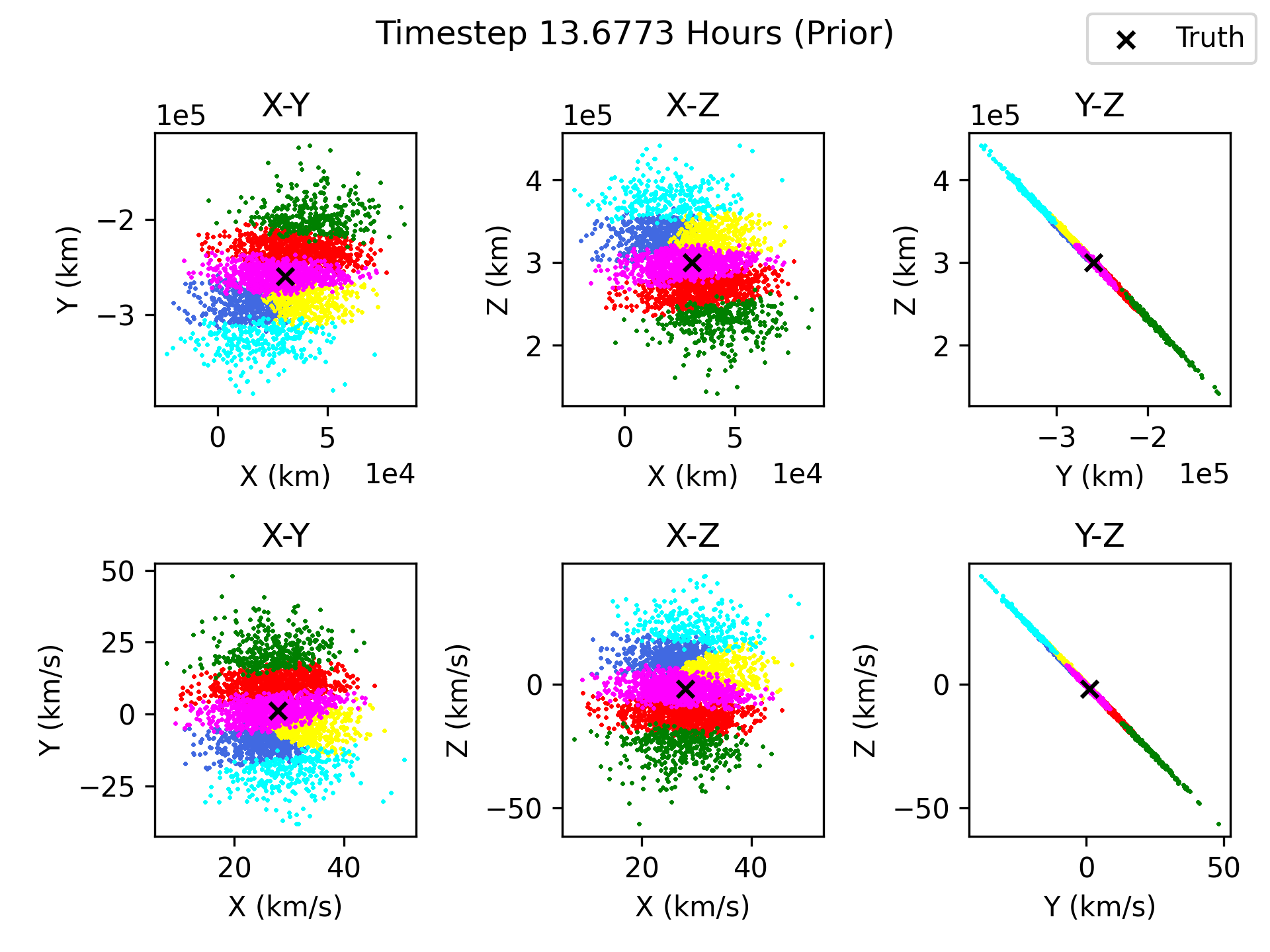}
        \caption{\textit{A priori} clustered estimate for a target at some point in time}
        \label{fig:2apgm2apriori}
    \end{subfigure}

    \vspace{0.2cm} 

    \begin{subfigure}{\textwidth}
        \centering
        \includegraphics[width=0.5\linewidth]{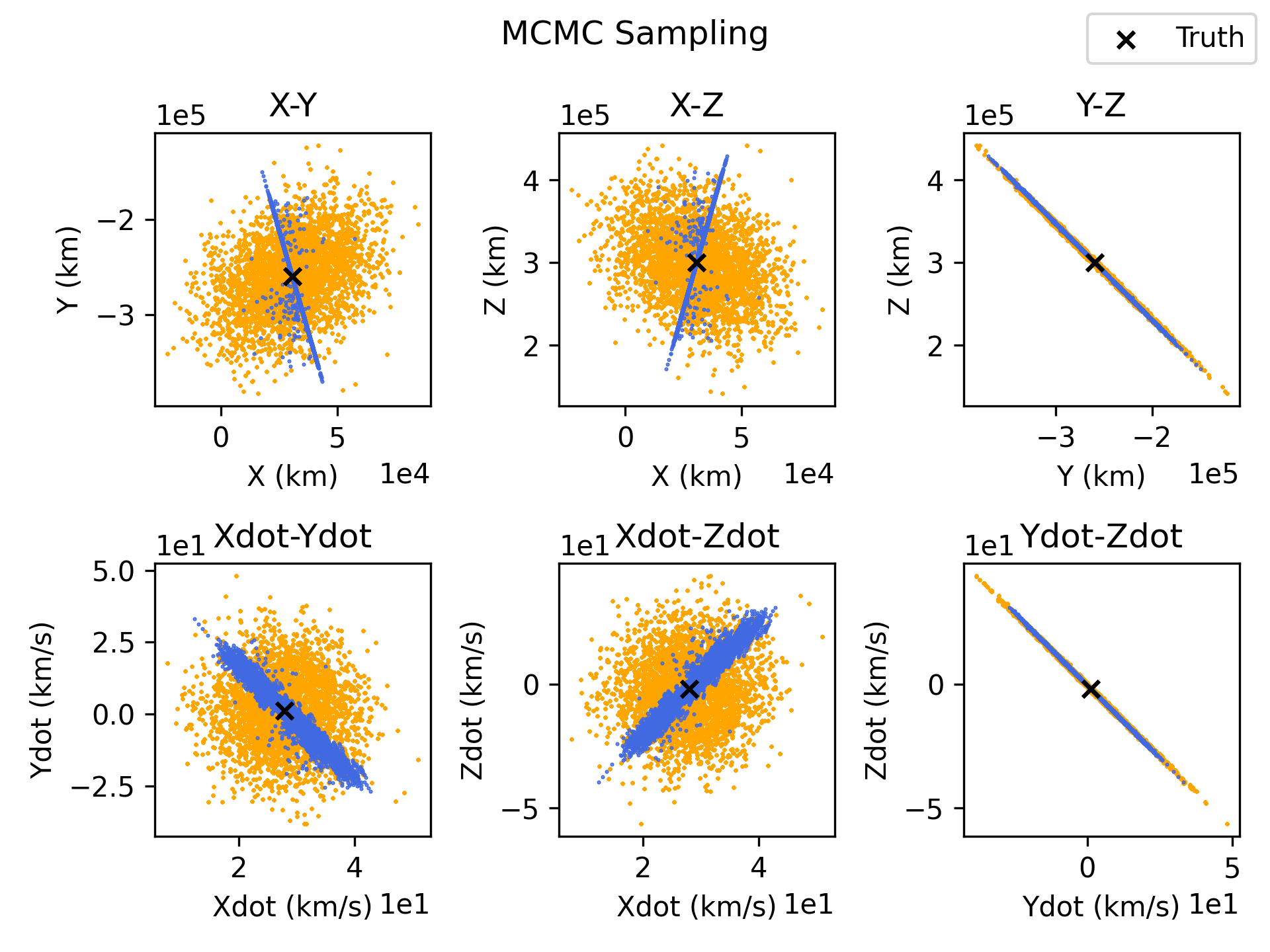}
        \caption{Collective MCMC chains $A_i$ for all $i$ after sample de-correlation}
        \label{fig:2bpgm2mcmcAll}
    \end{subfigure}

    \vspace{0.2cm} 

    \begin{subfigure}{\textwidth}
        \centering
        \includegraphics[width=0.5\linewidth]{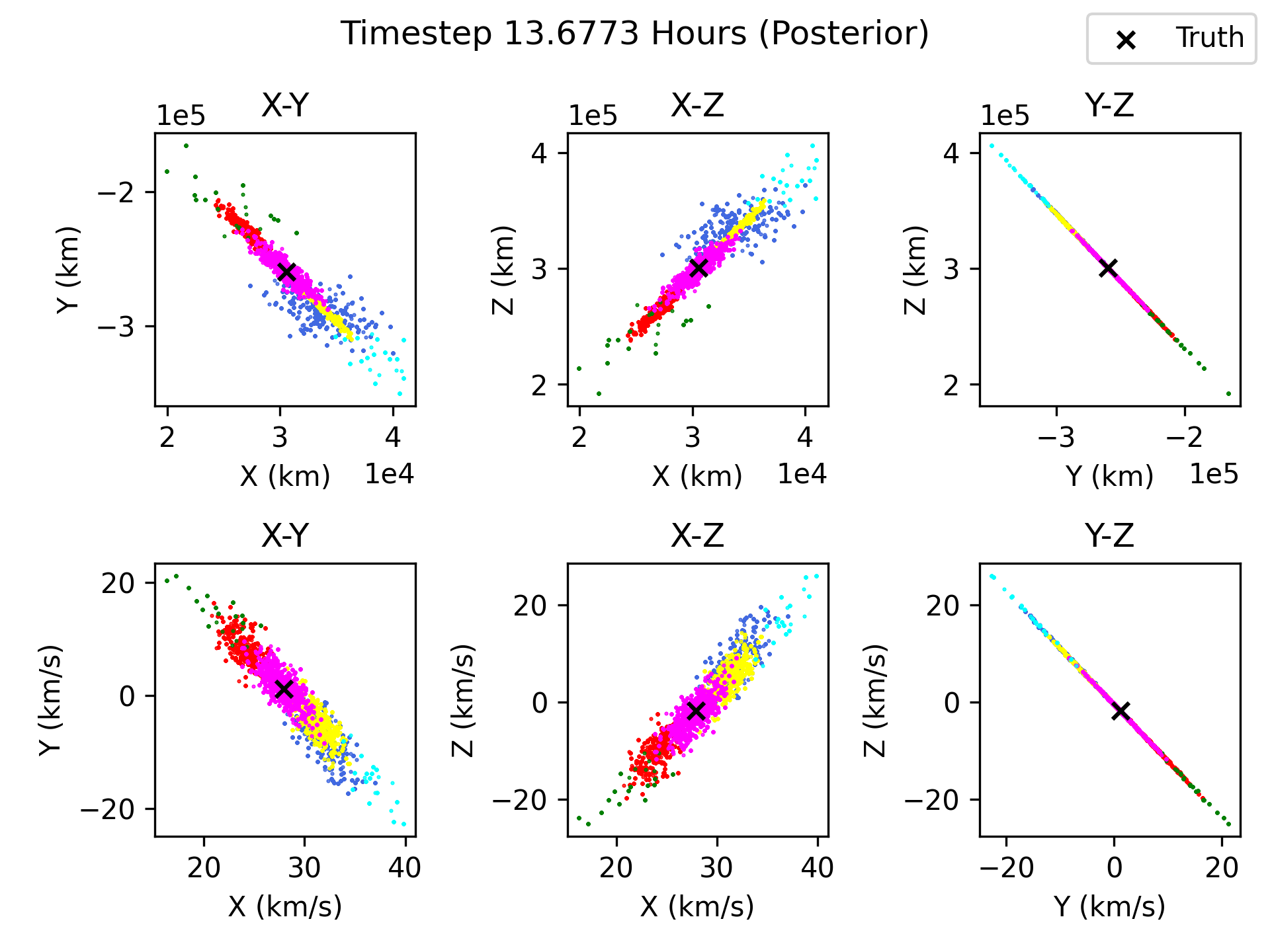}
        \caption{Posterior state estimate expressed as a GMM}
        \protect\label{fig:2cUpdate}
    \end{subfigure}

    \caption{PGM-II filtering MCMC sampling, clustering, and ensemble update steps outlining major parts of the larger update process. MCMC sampling narrows an \textit{a priori} estimate $p_k^{-}(\mathbf{x})$ into a narrow region of high likelihood given some $p(\mathbf{z}_k|\mathbf{x})$. MCMC samples are decorrelated into ensembles $A_i$ and clustered into posterior GMM components. An ensemble weight update is computed to obtain final posterior GMM component weights.}
    \label{fig:2pgm2demo}
\end{figure}

To summarize the PGM-II filter for a single iteration, we reuse the same resampling, propagation, and \textit{a priori} estimate clustering steps as the PGM-I filter. However, the update process of the second PGM filter is more computationally involved due to MCMC sampling and likelihood computation. We outline the PGM-II algorithm in Algorithm \ref{alg:pgm2}.\cite{raihan2018pgm2} We also illustrate the PGM-II update process in Figure \ref{fig:2pgm2demo}.

\begin{algorithm}
\caption{Particle Gaussian Mixture Filter II}\label{alg:pgm2}
\begin{algorithmic}
    \State Given $\pi_0 (x) = \sum_{i=1}^{M(0)} \omega_i (0) p_g(x; \mu_i (0), P_i (0))$, transition density kernel $p(x'|x)$, $n = 1$.
    \begin{enumerate}
        \State Sample $N_p$ particles from $\pi_{n-1}$ and the transition density kernel $p_n (x'|x)$ as follows: \label{alg:repeat2}
        \begin{itemize}
            \item Sample $X^{(i)^\prime}$ from $\pi_{n-1} (\cdot)$.
            \item Sample $X^{(i)}$ from $p(\cdot|X^{(i)^\prime})$.
        \end{itemize}
        \State Use a clustering algorithm $\mathcal{C}$ to cluster the set of particles $X^{(i)}$ into $M^{-} (n)$ Gaussian clusters with weights, mean, and covariance given by $\{\omega_i^{-} (n), \mu_i^{-} (n), P_i^{-} (n)\}$.
        \State Use MCMC to sample from the component posteriors $\pi_{i,n}(x)$ to generate the ensembles $A_i$.
        \State Compute the mixture weights $w_i(n)$ by evaluating the sequence of modal likelihoods using the likelihood function definition and the weight update equation.
        \State Sample N particles from the weighted collection of ensembles $\{w_i(n), A_{n,i}\}$.
        \State $n = n + 1$. Go to Step \ref{alg:repeat2}.
    \end{enumerate}
\end{algorithmic}\end{algorithm}

\section{Hybrid Particle Gaussian Mixture (H-PGM) Filtering Framework}\label{sec:hpgm}

In theory, all \textit{a priori} probability distributions may be expressed as a Gaussian Mixture Model. However, highly uninformative priors will require tens, or even hundreds, of GMM components, posing a dimensionality issue for both the Particle Gaussian Mixture filters. Two major limitations of both PGM filters is that both rely upon GMM-based \textit{a priori} estimates exclusively, and measurement noise has to be modeled as additive zero-mean Gaussian. The highest level of uncertainty is typically quantified by a uniform PDF. In this section, we describe a few extensions to the PGM-II filter that help us design a hybrid Particle Gaussian Mixture (H-PGM) filtering framework leveraging both PGM filters for a cislunar environment under extremely high state uncertainty.

\subsection{Extensions to the Particle Gaussian Mixture-II Filter}\label{subsec:extPGM2}

Due to a homogeneous probability density within a defined range, it is difficult to model any multivariate uniform PDF with a finite number of Gaussian Mixture Model components small enough to computationally justify MCMC sampling. As we shall demonstrate in Section \ref{subsec:limitations}, even if we are able to model a large multivariate uniform PDF as a Gaussian Mixture Model, the unwieldy uncertainty will eventually cause the PGM-I filter to lose target custody. We revisit the MCMC sampling function in Eq. \eqref{eq12:MCMCsamplingfunc}. We note that even though the PGM-II filter requires $p_k^{-}(\mathbf{x})$ and $p(\mathbf{z}_k|\mathbf{x})$ to be modeled as a GMM and as additive zero-mean Gaussian respectively. However, generalized MCMC samplers do not, and this allows us to generalize how target \textit{a priori} estimates and measurement noise and other fusion parameter PDFs are distributed. We test the H-PGM solution in unwieldy, high uncertainty environments by considering Baysian updates or target characteristic fusions in which either or both $p_k^{-}(\mathbf{x})$ and $p(\mathbf{z}_k|\mathbf{x})$ are modeled as (multivariate) uniform PDFs.

For the remainder of this work, we shall consider four types of information fusions related to Eq. \eqref{eq12:MCMCsamplingfunc}, placing heavy emphasis on the first three:
\begin{itemize}
    \item Gaussian/GMM-distributed $p_k^{-}(\mathbf{x})$ with Gaussian-distributed $p(\mathbf{z}_k|\mathbf{x})$
    \item Gaussian/GMM-distributed $p_k^{-}(\mathbf{x})$ with uniformly distributed $p(\mathbf{\Theta}_k|\mathbf{x})$
    \item Uniformly distributed $p_k^{-}(\mathbf{x})$ with Gaussian-distributed $p(\mathbf{z}_k|\mathbf{x})$
    \item Uniformly distributed $p_k^{-}(\mathbf{x})$ with uniformly distributed $p(\mathbf{\Theta}_k|\mathbf{x})$
\end{itemize}
The two traditional Particle Gaussian Mixture filters solely consider the first fusion. For the second and fourth fusions, we invoke the target characteristic vector notation in Section \ref{subsec:RSOchars}, noting that measurements are typically given as a single value with an associated noise distribution, whereas target \textit{a priori} information is given as a multivariate uniform PDF with heuristic constraints (e.g. telescope angular velocities are no greater than a certain value). We shall only consider the third fusion in this list for the very first time step, in which we start with a highly uncertain, multivariate uniform PDF $p_0^{-}(\mathbf{x})$. We describe how to obtain this $p_0^{-}(\mathbf{x})$, also known as our IOD estimate, in Section \ref{subsec:newIOD}.

\subsection{New Cislunar Initial Orbit Determination: Modeling Extreme Initial State Uncertainty} \label{subsec:newIOD}

As noted in Section \ref{sec:1intro}, literature about probabilistic IOD techniques for a cislunar environment is sparse. Many probabilistic target tracking problems start with a Gaussian distribution whose mean is centered at some point generated using a deterministic initial orbit determination method and whose covariance is small. Sometimes, these means are centered directly at the target truth. In reality, making a good deterministic initial state estimate requires many pieces of well-defined \textit{a priori} information. Furthermore, many filters are known to fail in the face of extreme initial state uncertainty. To that end, we develop an initial state estimate that leverages only two pieces of \textit{a priori} information: 1) the range within which the cislunar domain is defined, and 2) a sufficiently large target speed bound. 

For part of our KF-PGM work, we noted that we may be able to reasonably characterize cislunar observer-target ranges to lie between two times the geosynchronous (GEO) orbit distance and below roughly 550000 km.\cite{paranjape2026} This implies that the $x$, $y$, and $z$ coordinates of the state vector in the ECI reference frame are bounded by $\pm550000$ km. apiece. Letting $\mathbf{x}(j)$ represent the $j$-th index of the state vector in the ECI reference frame, we can mathematically define the uniform PDF for $1 \leq j \leq 3$ as $\mathbf{x}(j) \sim \mathcal{U} \lbrack -550000, 550000 \rbrack$ km. We note that the ranges encompassed by these uniform PDF bounds extend beyond the 550000 km. upper bound of cislunar space and also encompass regions defined by the two-body problem with the Earth. We nevertheless include these regions within our initial state estimate to emphasize that our hybrid PGM-based orbit determination framework is robust to highly uncertain position estimates.

\begin{figure}[h!]
	\centering\includegraphics[width=\columnwidth]{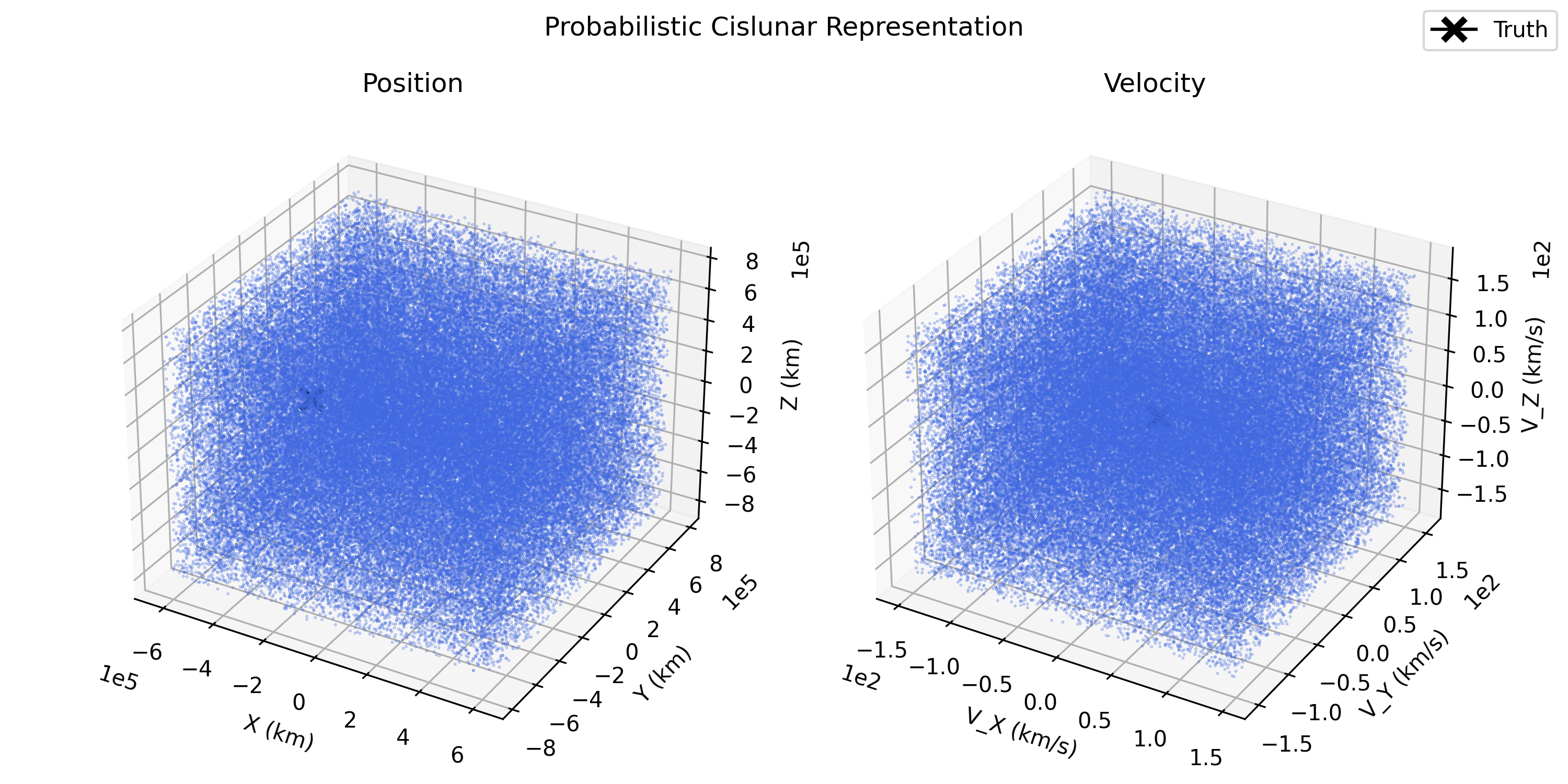}
	\caption{A large multivariate uniform PDF representing $p_0^{-}(\mathbf{x})$, our initial state estimate in the ECI reference frame. Samples within this multivariate uniform PDF may be converted to the synodic or topocentric frames to obtain appropriate uniform PDF bounds in those reference frames.}
	\label{fig:3mvUniform}
\end{figure}

In the same KF-PGM work, we introduced a valid assumption that cislunar \textit{resident} space objects are unlikely to escape the gravitational pull of the solar system and are therefore confined to speeds less than or equal to 42 km/s.\cite{paranjape2026} However, we extend that speed bound to 100 km/s to begin with a distribution whose spread can be characterized by an entropy greater than that produced by the minimal range assumption from our KF-IOD work.\cite{paranjape2026}. Similar to our position PDF formulation, we set  $\mathbf{x}(j) \sim \mathcal{U} \lbrack -100, 100 \rbrack$ km/s for $4 \leq j \leq 6$ (i.e. the velocity components). Together, these six restrictions form a PDF which we shall consider as the initial state estimate or the IOD equivalent of this work. Figure \ref{fig:3mvUniform} illustrates this large, multivariate normal PDF in the state space. Subsequent state updates and predictions are modeled as GMMs.

We note that the distribution in Figure \ref{fig:3mvUniform} may be used to initial cislunar RSOs states at any time step. While KF-PGM's IOD method has allowed us to develop more refined initial state estimates with a large, singular range bound, it requires several measurements for consistent state estimates over time. When we encounter infrequent observations, we no longer need to rely upon frequent observations to develop an IOD estimate; initialization with the PDF shown in Figure \ref{fig:3mvUniform} will suffice. In Section \ref{subsec:limitations}, we shall expand upon some of the limitations of the KF-PGM framework and how our H-PGM solution obviates the need for several short-arc observations. 

Compared to the standard PGM-II algorithm described in Section \ref{subsec:pgm2}, no major modifications are required to Algorithm \ref{alg:pgm2} to fuse the uniform PDF with observations whose noise is modeled as additive, zero-mean Gaussian. After the MCMC sampling step of Algorithm \ref{alg:pgm2}, de-correlating the resulting ensembles will involve modeling the resulting state estimate and any subsequent state estimates into Gaussian Mixture Models. As a result, the fusion of a uniformly distributed $p_k^{-}(\mathbf{x})$ with a Gaussian-distributed $p(\mathbf{z}_k|\mathbf{x}_k)$ will only need to be performed once.

\subsection{Information Fusion Leveraging Target Characteristics}\label{subsec:fuseWithTheta}

In Section \ref{subsec:RSOchars}, we introduced a variable $\mathbf{\Theta}_k$ to represent some target \textit{a priori} information at time step $k$. We make clear distinctions between $\mathbf{z}_k$ and $\mathbf{\Theta}_k$, namely that target characteristics are not considered part of the measurement model $\mathbf{h}(\mathbf{x}_k)$ and that measurement noise $p(\mathbf{z}_k|\mathbf{x})$ is distributed as Gaussian whereas $p(\mathbf{\Theta}_k|\mathbf{x})$ is distributed uniformly. Since measurement noise and target \textit{a priori} information are distributed differently, it helps to perform MCMC sampling sequentially. Expressing the resulting MCMC sampling function $P(\mathbf{x}_k)$ at time step $k$ when both target \textit{a priori} information and angles-only observations are available as
\begin{equation}\label{eq23:sequentialMCMCsamp}
        P(\mathbf{x}_k) = p^{-}(\mathbf{x}_k)p(\mathbf{\Theta}_k|\mathbf{x})p(\mathbf{z}_k|\mathbf{x}) = p^{-}(\mathbf{x}_k)p(\mathbf{z}_k|\mathbf{x})p(\mathbf{\Theta}_k|\mathbf{x}),
\end{equation}
we notice that we can either first fuse our target \textit{a priori} estimate with our target characteristic PDF and then our angles-only measurements or first fuse our target \textit{a priori} estimate with our angles-only measurements and then our target characteristic PDF. 

It is recommended to do MCMC sampling one product at a time since ensemble likelihoods are computed differently based on how $p(\mathbf{z}_k|\mathbf{x})$ and $p(\mathbf{\Theta}_k|\mathbf{x})$ are distributed. It may also be noted per Section \ref{subsec:newIOD} that when $k = 0$, the fusion of $p^{-}(\mathbf{x}_k)$ with $p(\mathbf{\Theta}_k|\mathbf{x})$ involves fusing two uniform PDFs, whose resulting representation we would model as a GMM. Even when $p(\mathbf{\Theta}_k|\mathbf{x})$ is uniformly distributed, the modified MCMC sampling and subsequent steps of Algorithm \ref{alg:pgm2} are similar to the original algorithm. However, likelihood computation and the subsequent ensemble weight update steps will differ.

We begin by observing that when $p(\mathbf{\Theta}_k|\mathbf{x})$ is uniformly distributed and normalized, the conditional PDF may be expressed as an indicator function $I_F(\mathbf{x}_k)$, which defines whether or not some state estimate $\mathbf{x}_k$ drawn from $p_k^{-}(\mathbf{x})$ lies within the target characteristic bounds. Bayes' rule and the law of total probability then give us 
\begin{equation}\label{eq24:if_bayes}
        p(\mathbf{x}|\mathbf{\Theta}_k) = \frac{p_k^{-}(\mathbf{x})p(\mathbf{\Theta}_k|\mathbf{x})}{\int p_k^{-}(\mathbf{x})p(\mathbf{\Theta}_k|\mathbf{x}) d\mathbf{x}} = \frac{p_k^{-}(\mathbf{x})I_F(\mathbf{x})}{\int p_k^{-}(\mathbf{x})I_F(\mathbf{x}) d\mathbf{x}}.
\end{equation}
Noting that the likelihood function may be expressed as $l_i(\mathbf{\Theta}_k) = \int p_k^{-}(\mathbf{x})I_F(\mathbf{x}) d\mathbf{x}$, we derive the following approximation:
\begin{equation}\label{eq25:uniformLikelihood}
       l_i(\mathbf{\Theta}_k) = \int p_k^{-}(\mathbf{x})I_F(\mathbf{x}) d\mathbf{x} = \mathbb{E}[I_F(\mathbf{x})] \approx \frac{1}{S_i}\sum_{s=1}^{S_i} I_F(\mathbf{x}).
\end{equation}
In other words, given an ensemble of de-correlated, MCMC-sampled states $A_i$, the likelihood is simply the number of particles consistent with the target \textit{a priori} information function $\mathbf{\Theta}_k = \mathbf{\theta}(\mathbf{x})$ over the total number of particles in $A_i$. Now that we show how to extend the PGM-II filter MCMC sampling, clustering/de-correlation, and the subsequent weight update, we can apply this extended PGM-II filter to subsequent state estimates.

\subsection{The Hybrid Solution}

In this section, we have so far discussed the four types of information fusions for $P(\mathbf{x})$ in Eq. \eqref{eq12:MCMCsamplingfunc}, a new uninformative but consistent IOD estimate for cislunar RSOs, and an extension of the PGM-II filter update process when $p(\mathbf{z}_k|\mathbf{x})$ or $p(\mathbf{\Theta}_k|\mathbf{x})$ are distributed differently and given at the same time. While the PGM-II filter has potential in a cislunar environment, its MCMC sampling step will have limited parallelization, resulting in a computationally exhaustive algorithm. For this reason, we recommend using a sequential mixture or \textit{hybrid} of the two PGM filters.

Due to high computational runtime of a single iteration of the PGM-II filter, it is recommended to run the filter with only the first $n_k$ observations. We recommend setting $n_k = 2$ since two spaced sets of angles-only measurements implicitly localize both position and velocity, whereas a single set of angles-only measurements only localizes the position. Once the PGM-II filter sufficiently localizes state estimates in the position and velocity spaces, it becomes more computationally efficient to switch to the PGM-I filter for subsequent target tracking. This hybrid solution is encapsulated in Algorithm \ref{alg:hybrid}.

\begin{algorithm}
\caption{Hybrid Particle Gaussian Mixture (H-PGM) Solution}\label{alg:hybrid}
\begin{algorithmic}
    \State Given a uniformly distributed initial PDF $p_0^{-}(\mathbf{x})$, transition density kernel $p(\mathbf{x}'|\mathbf{x})$, $n = 1$, switching time step $n_k$.
    \begin{enumerate}
        \State Sample $N_p$ particles from $\pi_{n-1}$ and the transition density kernel $p_n (x'|x)$ as follows: \label{alg:repeatHybrid}
        \begin{itemize}
            \item Sample $X^{(i)^\prime}$ from $\pi_{n-1} (.)$.
            \item Sample $X^{(i)}$ from $p(.|X^{(i)^\prime})$.
        \end{itemize}
        \State Use a clustering algorithm $\mathcal{C}$ to cluster the set of particles $X^{(i)}$ into $M^{-} (n)$ Gaussian clusters with weights, mean, and covariance given by $\{\omega_i^{-} (n), \mu_i^{-} (n), P_i^{-} (n)\}$.

        \If{$n \leq n_k$}
            \State Follow Steps 3-5 from Algorithm \ref{alg:pgm2}.
        \Else
            \State Follow Step 3 from Algorithm \ref{alg:pgm1}.
        \EndIf

        \State $n = n + 1$. Go to Step \ref{alg:repeatHybrid}.
    \end{enumerate}
\end{algorithmic}\end{algorithm}

\section{Results and Discussion}\label{sec:Results}

In this section, we demonstrate our H-PGM solution upon several orbits. We make several comparisons of this OD framework to the KF-PGM framework. Finally, we demonstrate the short- and long-term effects of several target characteristic fusions on target estimates. For each result provided in this section, we set our starting distribution $p_0^{-}(\mathbf{x})$ equal to the large, multivariate uniform PDF provided in Figure \ref{fig:3mvUniform}. The highly uncertain uniform PDF bounds serve to underscore the robustness of our H-PGM solution under minimal information. For each of the results discussed below, we assume that measurement noise $p(\mathbf{z}_k|\mathbf{x})$ is additive zero-mean Gaussian. We also assume that our observer is based in College Station, TX, USA.

\subsection{Important Noise Characteristics and Metrics}\label{subsec:noiseChars}

Since we assume that measurement noise $p(\mathbf{z}_k|\mathbf{x})$ is additive zero-mean Gaussian, we express its noise covariance matrix $\mathbf{R}$ as
\begin{equation}\label{eq26:OD_R}
    \mathbf{R} =
    \begin{bmatrix}
        (1.5 \space \text{arcsec})^2 & 0\\
        0 & (1.5 \space \text{arcsec})^2
    \end{bmatrix}.
\end{equation}
We chose an angular standard deviation of 1.5 arcsec due to the high precision of modern optical telescopes.\cite{mishra2024} We analyze the performance of the H-PGM solution, especially with respect to the KF-PGM framework and other filters, using the Shannon entropy metric $H(k)$ for some time step $k$. Abusing notation slightly, we notice that when $p(\mathbf{x})$ is a Gaussian Mixture Model, there is no closed-form solution to the Shannon entropy integral, given by 
\begin{equation}\label{eq27:entropyContinuous}
    H(k) = - \int_{\mathbb{R}^{n_x}} p(\mathbf{x}) \log{p(\mathbf{x})} d\mathbf{x}.
\end{equation}
As a result, we make a discrete approximation to the entropy $H(k)$ with
\begin{equation}\label{eq28:mcEntropy}
    H(k) = \mathbb{E}[- \log{p_{k}(\mathbf{x}_i)}] \approx -\frac{1}{N} \sum_{i=1}^{N} \log{p_k(\mathbf{x}_i)},
\end{equation}
where $p_k(\mathbf{x}_i)$ is the GMM-based probability density of some particle $\mathbf{x}_i$ at time step $k$ in our $N$-particle ensemble. For UKFs and Ensemble Kalman Filters (EnKFs), the closed-form integral solution in Eq. \eqref{eq27:entropyContinuous} is used.

\subsection{Example 1: 9:2 Resonant Near-Rectilinear Halo Orbit Baseline Solution}\label{subsec:ex1}

NASA's now-suspended Gateway program chose the 9:2 resonant near-rectilinear halo orbit (NRHO) based around the $L_2$ Lagrange point due to strong communication coverage of the lunar South Pole and low $\Delta v$ requirements for potential targets operating within the vicinity.\cite{2019NASA} Since this orbit has historically provided several promising results in cislunar target tracking literature, we shall demonstrate several examples of our H-PGM solution for a target operating in this orbit. Here, we demonstrate what we shall refer to as our \say{Baseline} solution. For the \say{Baseline} solution, we ignore target \textit{a priori} information from Section \ref{subsec:RSOchars}. Instead, we simply perform measurement updates using our extended PGM-II filter for the first two measurements before subsequently switching to the PGM-I filter. 

\begin{figure}[!thpb]
    \centering
    \begin{subfigure}{0.5\columnwidth}
        \centering
        \includegraphics[width=\linewidth]{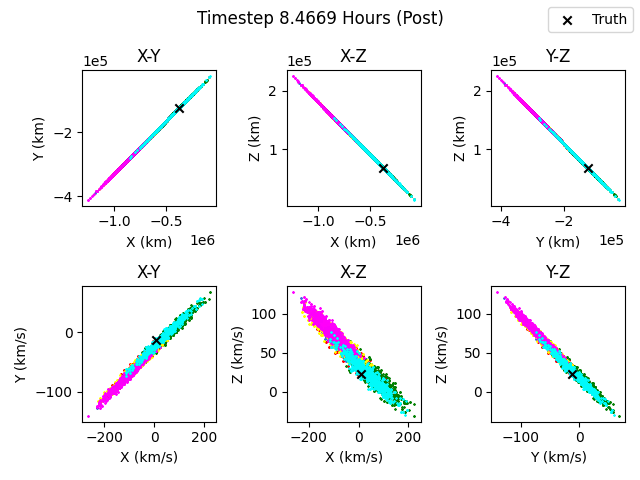}
        \caption{Posterior target estimate after two iterations of the modified PGM-II filter. Two iterations are necessary for sufficient, \say{pencil-like} localization in the state and velocity spaces.}
        \label{fig:4aPGM2}
    \end{subfigure}

    \vspace{0.2cm} 

    \begin{subfigure}{0.5\columnwidth}
        \centering
        \includegraphics[width=\linewidth]{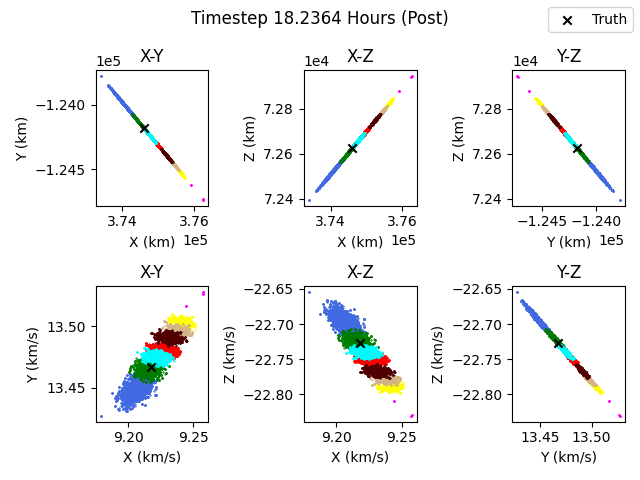}
        \caption{Posterior state estimate at the end of target's pass above the local horizon}
        \label{fig:4bHybrid}
    \end{subfigure}

    \caption{Target estimates resulting from H-PGM sequencing -- two iterations of PGM-II and then PGM-I thereafter -- for a target (truth denoted by a black \textbf{X}) within the 9:2 resonant NRHO.}
    \label{fig:4baselineFusions}
\end{figure}

For sensor tasking, we make our observer take measurements roughly every 40 minutes during the target's first pass, treating the target as a new object in the cislunar space object catalog. This measurement schedule is identical to the first pass measurement schedule described in the original KF-PGM works.\cite{paranjape2025, paranjape2026}. The target stays above the local horizon -- during which period our observer may search for the target -- for roughly 10 hours, after which the simulation ends. We choose a short time period in this example to focus on the short-term benefits of the H-PGM solution. We begin with the IOD estimate shown in Figure \ref{fig:3mvUniform}. We showcase the posterior estimates at the end of the second time step -- when we switch filters -- and at the end of the first pass in Figure \ref{fig:4baselineFusions}.

Observing the axis labels between Figures \ref{fig:3mvUniform} and \ref{fig:4bHybrid}, we see that at the end of the simulation, the position estimate is improved from the hundreds of thousands of kilometers to the \textit{tens} of kilometers and the velocity estimate is improved from hundreds of kilometers per second to tens of \textit{meters} per second. With relatively frequent observations, our H-PGM solution significantly reduces target uncertainty without being overconfident. To compare the H-PGM solution with the KF-PGM framework, we compare component state estimate precisions at the IOD stage and the end of the simulation with KF-PGM. 

\begin{figure}[!thpb]
    \centering
    \begin{subfigure}{0.5\columnwidth}
        \centering
        \includegraphics[width=\linewidth]{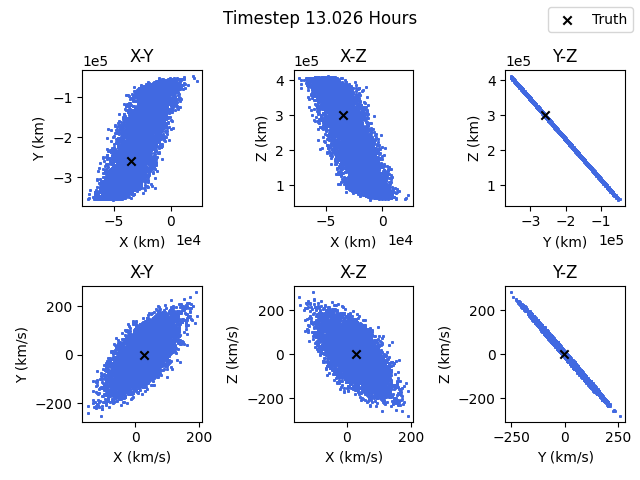}
        \caption{IOD estimate resulting from kinematically fitting polynomials through the first eight observations of the first pass along with a uniform PDF defining the range bounds of the cislunar domain.}
        \label{fig:5aIODcloud}
    \end{subfigure}

    \vspace{0.2cm} 

    \begin{subfigure}{0.5\columnwidth}
        \centering
        \includegraphics[width=\linewidth]{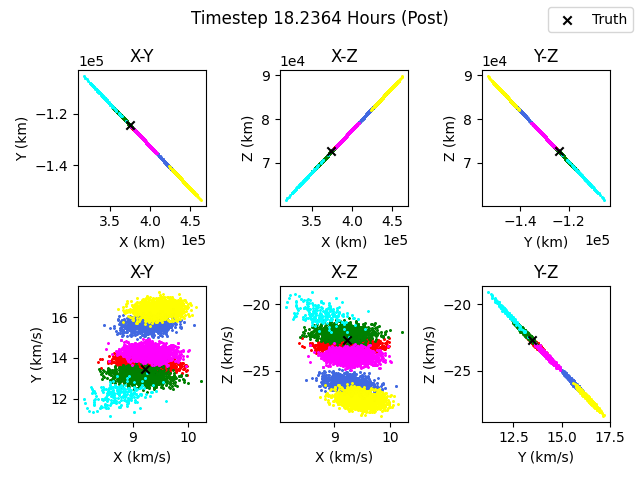}
        \caption{Posterior state estimate at the end of target's pass above the local horizon, leveraging eight observations}
        \label{fig:5bEoS}
    \end{subfigure}

    \caption{Target state estimates resulting from a kinematic fitting IOD procedure and PGM-I filtering through the end of one pass. The same measurements and schedule as Figure \ref{fig:4baselineFusions} are utilized.}
    \label{fig:5kfPGMExample}
\end{figure}

The probabilistic IOD aspect of KF-PGM (i.e. the \say{KF} part) involves the fusion of several consecutive $[AZ, EL]^T$ measurement sets with a uniform PDF defining the possible ranges of cislunar space. Leveraging measurement noise statistics and the range PDF, we draw several i.i.d samples of the three quantities to generate several initial position estimates, through which we fit polynomials strictly as functions of time. We take the derivatives of these polynomials to obtain the resulting velocity estimates at the last fitted observation's time step. The result is a large particle cloud such as that shown in Figure \ref{fig:5kfPGMExample}. Since this state estimate has some localization in position and velocity (especially for the $Y$-$Z$ and $\dot{Y}$-$\dot{Z}$ projections), we utilize the PGM-I filter for uncertainty reduction. We provide plots of the initial and final time steps, as well as a table of standard deviation values of each state vector component in Figure \ref{fig:5kfPGMExample} and Table \ref{table:stDevs}, respectively.

An important distinction between the KF-PGM framework and our H-PGM solution is the number of observations utilized for orbit determination (OD) strictly. For KF-PGM, instead of placing assumptions on valid cislunar target states, we abstract valid observer-target ranges into a PDF and fit a state estimate through the first $M$ measurements. While $M\geq2$ in theory, as we shall demonstrate in Section \ref{subsec:limitations}, we need a sufficiently large number of measurements through which to fit polynomials to mitigate PGM-I filter overconfidence. For Figure \ref{fig:5kfPGMExample}, $M = 8$, corresponding to half the measurements within the first pass.

\begin{table}[h]
    \caption{Target estimate standard deviations for KF-PGM and the H-PGM solution at the IOD stage and at the end of the first pass}
    \label{table:stDevs}
    \begin{center}
        \begin{tabular}{|c||c||c||c||c|}
        \hline
        Component & KF-PGM Start & KF-PGM End & H-PGM Start & H-PGM End \\
        \hline
        $x$ (km)      & 15370   & 29869   & 359028 & 406.47 \\ \hline
        $y$ (km)      & 82919   & 9902.4  & 444363 & 135.18  \\ \hline
        $z$ (km)      & 95764   & 5791.4  & 449691 & 77.709  \\ \hline
        $\dot{x}$ (km/s) & 54.047  & 0.32216 & 85.836 & 0.010630 
        \\ \hline
        $\dot{y}$ (km/s) & 78.993  & 1.2984  & 88.528 & 0.014715 
        \\ \hline
        $\dot{z}$ (km/s) & 88.644  & 1.9414  & 97.135 & 0.025009 
        \\ \hline
        \end{tabular}
    \end{center}
\end{table}

\subsection{Example 2: $L_2$ Lagrange Point Target Trajectory Solution}\label{subsec:ex2}

In some of our KF-PGM based works, we highlighted the importance of the first three Lagrange points within CR3BP dynamics.\cite{paranjape2025, paranjape2026} We specifically pointed out how the $L_1$, $L_2$, and $L_3$ points are points of unstable equilibrium, and how localized distributions propagated through this point experience some of the most chaotic warping or bifurcations. In particular, we presented a trajectory in which Gaussian-like \textit{a priori} estimates warp and bifurcate, but are re-localized efficiently with the PGM-I filter. We apply our H-PGM solution on this trajectory.

\begin{figure}[!thpb]
    \centering
    \begin{subfigure}{0.5\columnwidth}
        \centering
        \includegraphics[width=\linewidth]{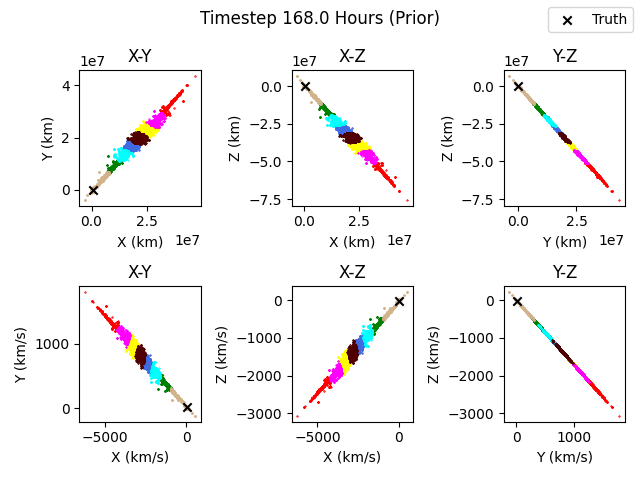}
        \caption{\textit{A priori} estimate of a target after it has passed through the $L_2$ Lagrange point and has reappeared in the sensor FOV.}
        \label{fig:6aAPriori}
    \end{subfigure}

    \vspace{0.2cm} 

    \begin{subfigure}{0.5\columnwidth}
        \centering
        \includegraphics[width=\linewidth]{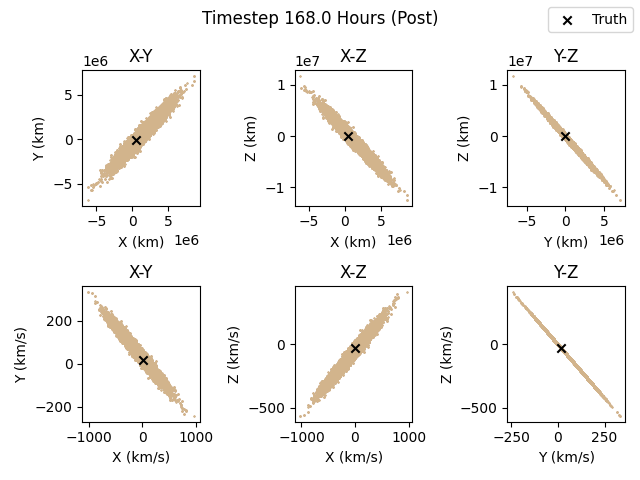}
        \caption{Posterior state estimate of the target in Figure \ref{fig:6aAPriori} after applying the PGM-I filter update step.}
        \label{fig:6bPosterior}
    \end{subfigure}

    \caption{State estimates of a target passing through the $L_2$ Lagrange point behind the Moon after it reappears within our sensor FOV. The H-PGM solution is applied to this trajectory.}
    \label{fig:6L2Example}
\end{figure}

\begin{figure}[!thpb]
    \centering
    \begin{subfigure}{0.5\columnwidth}
        \centering
        \includegraphics[width=\linewidth]{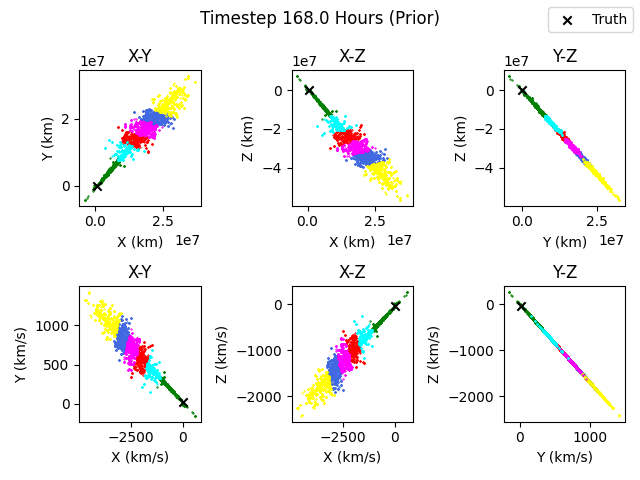}
        \caption{\textit{A priori} estimate of a target after it has passed through the $L_2$ Lagrange point and has reappeared in the sensor FOV.}
        \label{fig:7aAPriori}
    \end{subfigure}

    \vspace{0.2cm} 

    \begin{subfigure}{0.5\columnwidth}
        \centering
        \includegraphics[width=\linewidth]{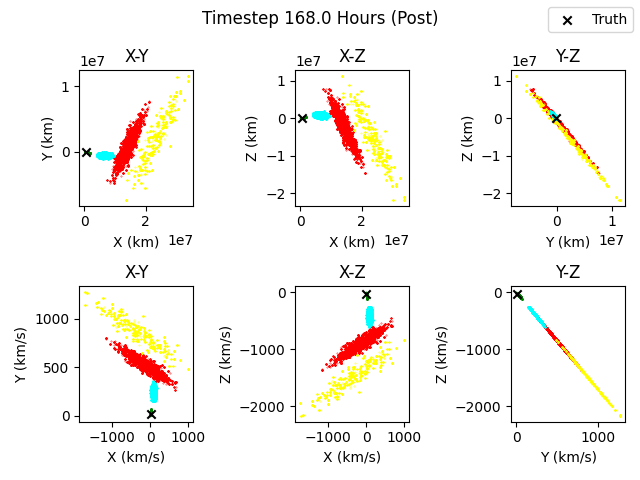}
        \caption{Posterior state estimate of the target in Figure \ref{fig:6aAPriori} after applying the PGM-II filter update step.}
        \label{fig:7bPosterior}
    \end{subfigure}

    \caption{State estimates of a target passing through the $L_2$ Lagrange point behind the Moon after it reappears within our sensor FOV using solely the extended PGM-II filter.}
    \label{fig:7L2Example_PGM2}
\end{figure}

\begin{figure}[!thpb]
    \centering
    \begin{subfigure}{0.5\columnwidth}
        \centering
        \includegraphics[width=\linewidth]{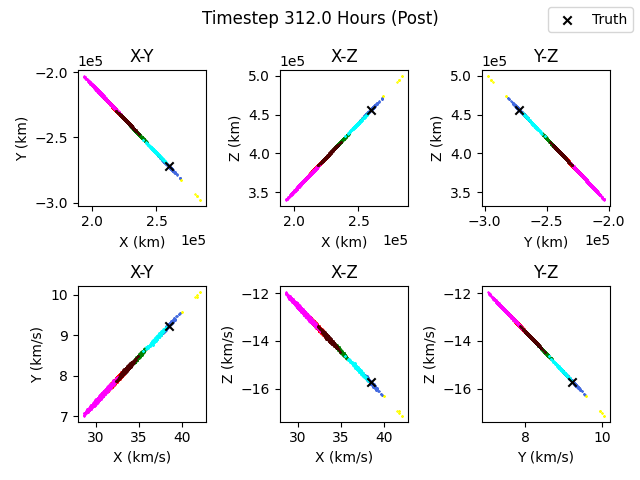}
        \caption{Final target estimate resulting from our hybrid PGM-based filtering solution.}
        \label{fig:8aHybrid}
    \end{subfigure}

    \vspace{0.2cm} 

    \begin{subfigure}{0.5\columnwidth}
        \centering
        \includegraphics[width=\linewidth]{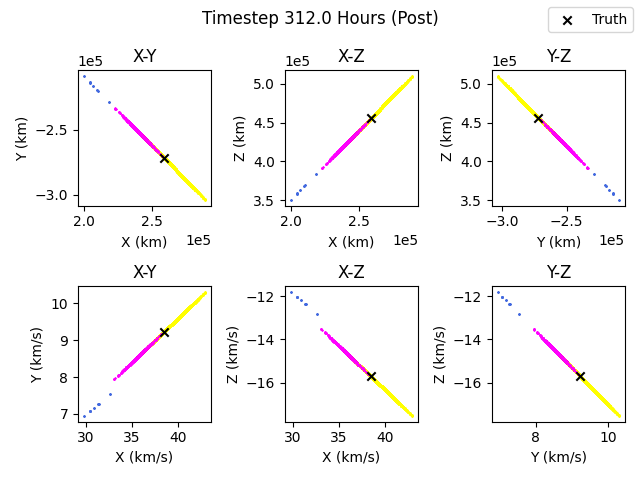}
        \caption{Final target estimate resulting from a solely PGM-II based filtering solution.}
        \label{fig:8bPGM2}
    \end{subfigure}

    \caption{Final state estimates of a target passing through the $L_2$ Lagrange point using the H-PGM and extended PGM-II filtering methods.}
    \label{fig:8finalL2Estimates}
\end{figure}

To demonstrate that our H-PGM solution is effective for sparser observations, we schedule measurements every 24 hours. Because the $L_2$ Lagrange point lies behind the Moon, there is roughly a five-day period during which no optical telescope can observe the target state. By that point, our hybrid PGM filtering method has already transitioned to the PGM-I filter. In Figure \ref{fig:6L2Example}, we illustrate the estimates just before and after an observation update after that five-day period with no observations. 
When we apply the extended PGM-II filter exclusively to this same trajectory for the same set of observations (i.e. without switching to the PGM-I filter), we shall notice in Figure \ref{fig:7L2Example_PGM2} vastly differing posterior state estimates once the target re-emerges from behind the Moon. 
Unlike the PGM-I filter update step, which places heavier emphasis on measurement likelihoods, multiple clusters become consistent with the PGM-II update, resulting in a multi-modal distribution for proceeding time steps. However, this multi-modality is temporary and disappears after five days of observations. Figure \ref{fig:8finalL2Estimates} depicts the posterior estimates 12 days after the target passes through the $L_2$ Lagrange point using the H-PGM and extended PGM-II solutions. These final state estimates become roughly as precise as each other.

\subsection{Example 3: Southern Butterfly Orbit H-PGM Solution}\label{subsec:ex3}

In Sections \ref{subsec:ex1} and \ref{subsec:ex2}, we demonstrated the effectiveness of our H-PGM solution for two different trajectories, including when measurements are both closely-spaced and infrequent. Furthermore, in Section \ref{subsec:ex2}, we demonstrated the effectiveness of the extended PGM-II filtering solution. In this subsection, we shall discuss the effectiveness of the H-PGM solution over the extended PGM-II filtering method for the Southern Butterfly Orbit. We chose this orbit due to its relatively short period (roughly 12 days) and bifurcation from the NRHO family of orbits.\cite{spreen2017, zimovan2017}

\begin{figure}[!thpb]
    \centering
    \begin{subfigure}{0.5\columnwidth}
        \centering
        \includegraphics[width=\linewidth]{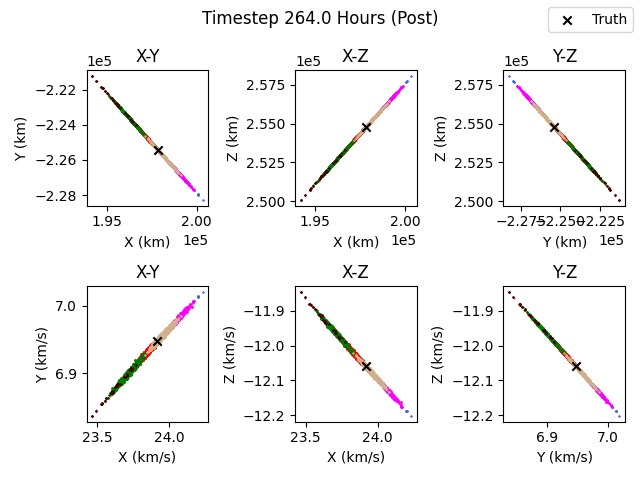}
        \caption{Final target estimate resulting from our hybrid PGM-based (H-PGM) filtering solution.}
        \label{fig:9aHybrid}
    \end{subfigure}

    \vspace{0.2cm} 

    \begin{subfigure}{0.5\columnwidth}
        \centering
        \includegraphics[width=\linewidth]{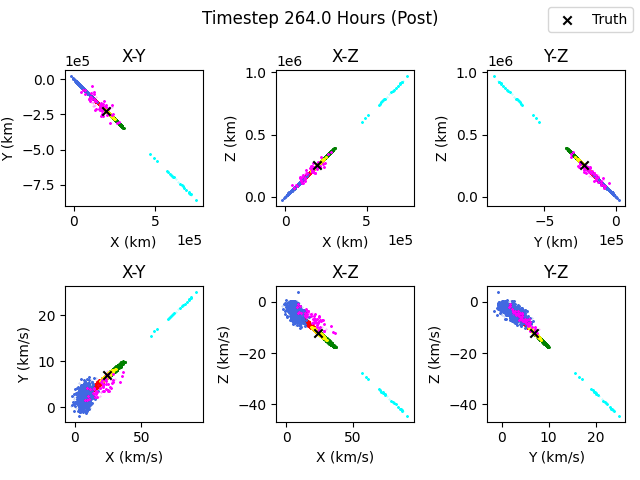}
        \caption{Final target estimate resulting from the extended PGM-II filtering solution.}
        \label{fig:9bPGM2}
    \end{subfigure}

    \caption{Final state estimates of a target within the Southern Butterfly Orbit after one full orbit with the H-PGM and extended PGM-II solutions.}
    \label{fig:9finalSBOEstimates}
\end{figure}

\begin{figure}[h!]
	\centering\includegraphics[width=0.33\columnwidth]{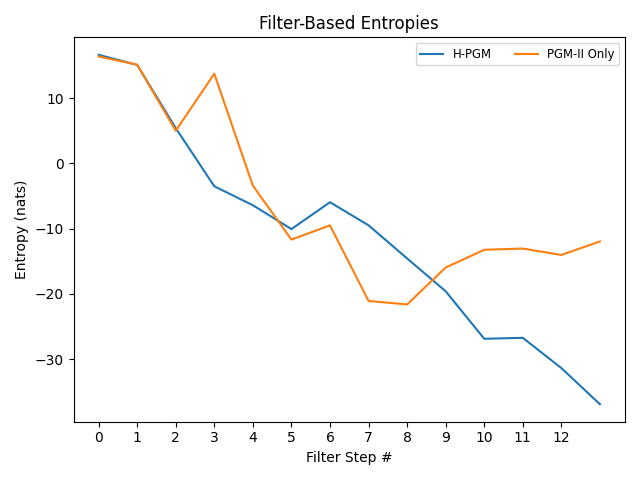}
	\caption{A comparison of the entropy evolution of the state estimates of a target within the Southern Butterfly Orbit with the hybrid PGM-II (H-PGM) filtering method and a purely recursive extended PGM-II method.}
	\label{fig:10SBOEntropies}
\end{figure}

For the first pass of this target trajectory, we schedule measurements every three hours to allow for some localization with the extended PGM-II filter. For each pass thereafter, we attempt to take a single observation at a random point in the pass. Starting with the same initial state estimate shown in Figure \ref{fig:3mvUniform}, we arrive at two different posterior target estimates in Figure \ref{fig:9finalSBOEstimates} at the end of one full orbit with the H-PGM and extended PGM-II filtering solutions.
To better quantify the effectiveness of the H-PGM solution over a purely recursive extended PGM-II solution for the Southern Butterfly Orbit, we utilize the entropy metric from Eq. \eqref{eq28:mcEntropy} for Figure \ref{fig:10SBOEntropies}.

As discussed in Section \ref{subsec:ex2}, the PGM-I filter update step tends to be more effective about eliminating the number of clusters especially when measurements are sparse. Since measurements for this orbit become more random and sparse beyond a certain point, the PGM-II filter tends to retain several clusters and distributions from its \textit{a priori} estimates, resulting in larger, more multi-modal posterior distributions. As a result, the effectiveness of the PGM-II filter reduces considerably, while the PGM-I filter continues to robustly estimate the target state more confidently. 

\subsection{Limitations of KF-PGM and Exclusive PGM-I Based Approaches}\label{subsec:limitations}

In Sections \ref{subsec:ex1} through \ref{subsec:ex3}, we demonstrated the ability of our H-PGM solution on a variety of chaotic orbits and trajectories in cislunar space. We also highlighted the H-PGM solution's superiority over the KF-PGM solution's over the same measurement set in \ref{table:stDevs} due to the H-PGM solution's observation-efficient approach. Although exclusive use of the PGM-I filter has been highlighted and demonstrated in several space domain awareness works, discussion of the limitations of PGM-I based approaches has been sparse.\cite{bolden2022, griggs2023, mishra2024, paranjape2025, paranjape2026} We include this subsection discuss some limitations of KF-PGM and PGM-I based approaches to further highlight the need of our H-PGM solution.

\begin{figure}[!thpb]
    \centering
    \begin{subfigure}[t]{0.45\textwidth}
        \centering
        \includegraphics[width=\linewidth]{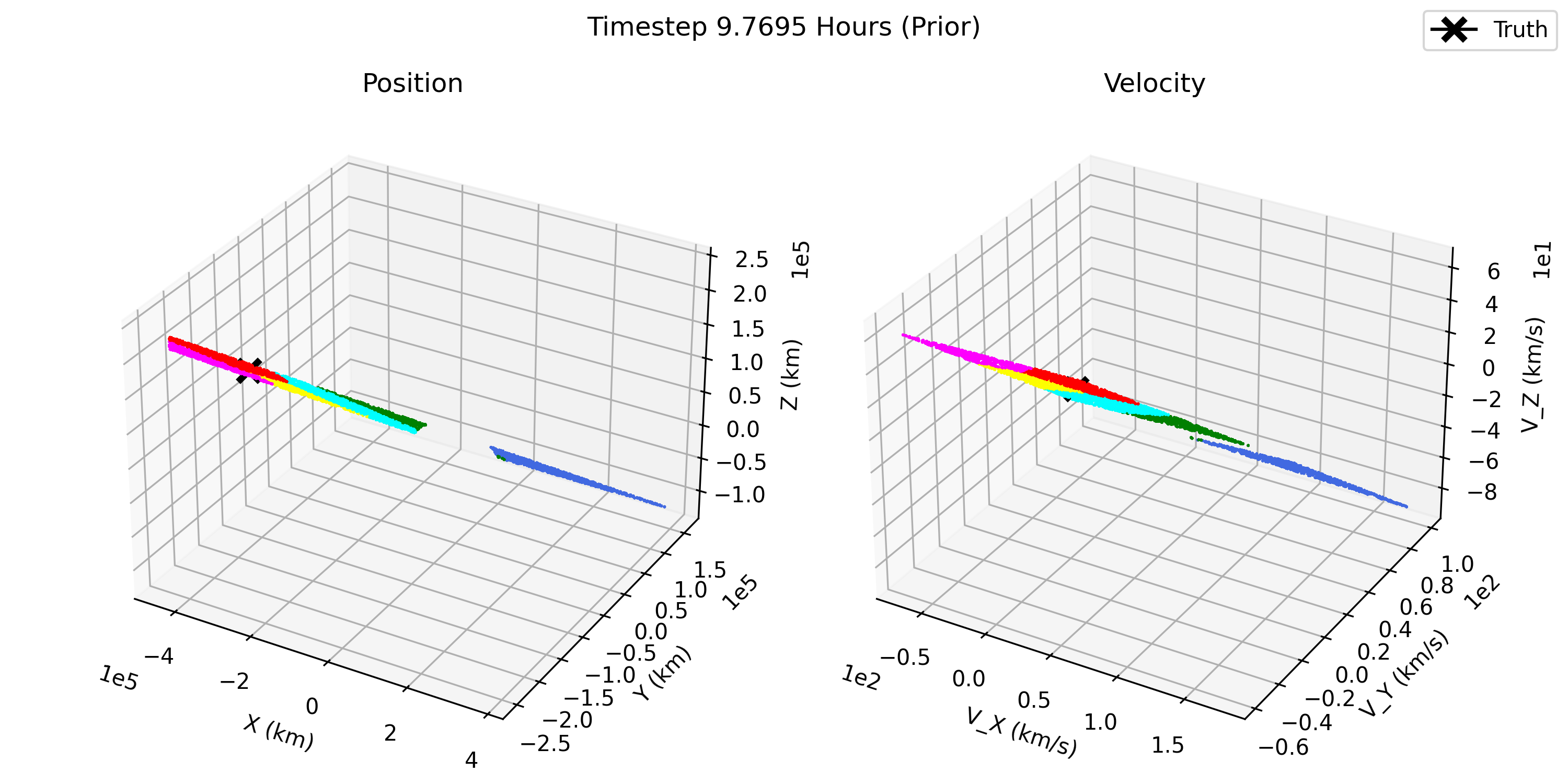}
        \caption{\textit{A priori} estimate propagated and clustered from an IOD estimate formulated from a linear kinematic fit}
        \label{fig:11aPrior}
    \end{subfigure}
    
    \begin{subfigure}[t]{0.45\textwidth}
        \centering
        \includegraphics[width=\linewidth]{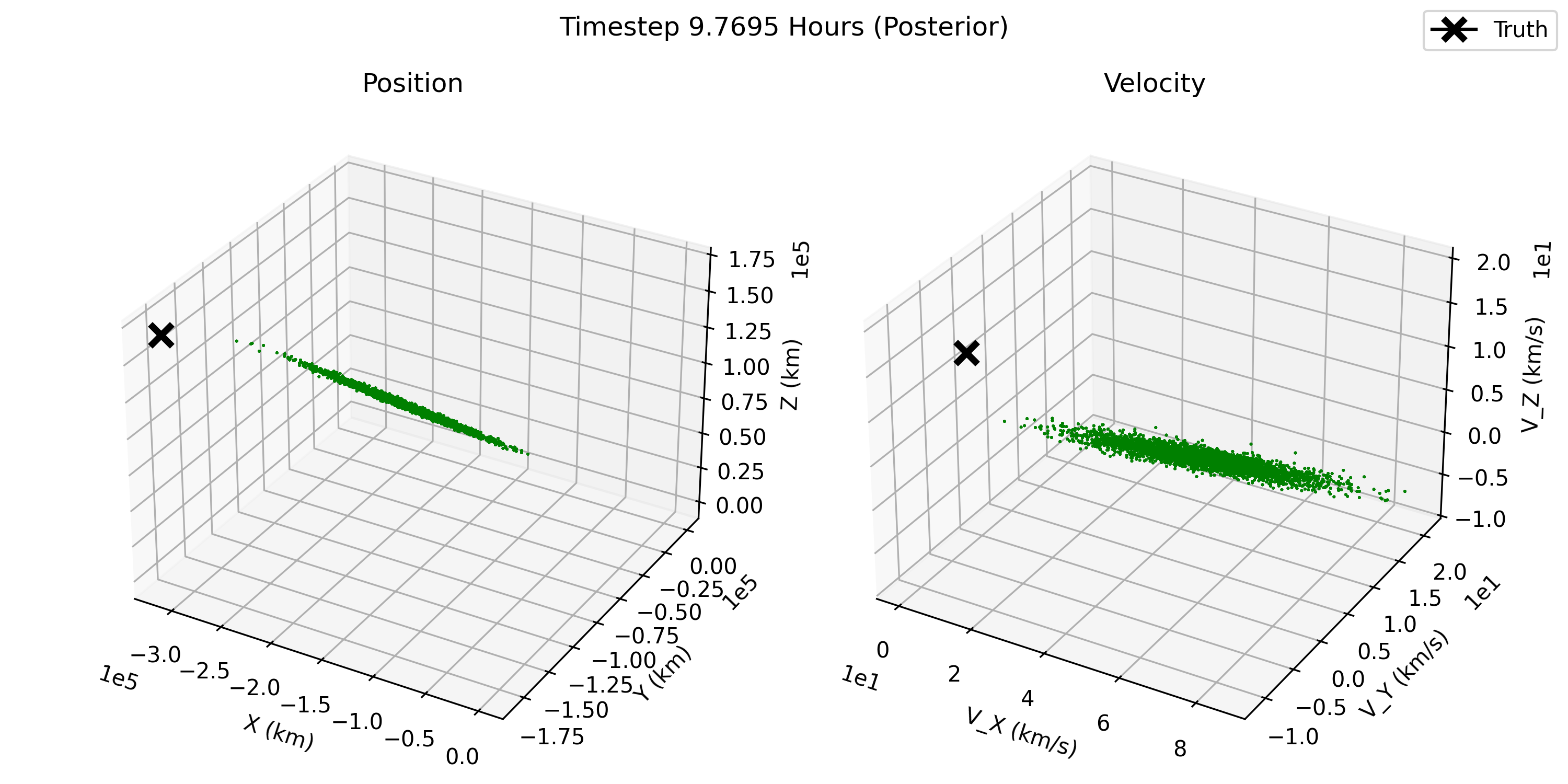}
        \caption{An inconsistent posterior estimate after application of a single PGM-I update step}
        \label{fig:11bPosterior}
    \end{subfigure}

    \caption{Limitations of utilizing KF-PGM with a linear fit between the first two measurements}
    \label{fig:11kfpgmfail}
\end{figure}

The KF-PGM approach requires polynomial fitting over some measurement set extrapolated several thousands of times with some range information and measurement noise. In theory, the minimum number of measurements through which we need to do kinematic fitting is two (i.e. a linear fit), so that a constant time derivative for the velocity exists. While this approach yields a consistent initial state estimate, the PGM-I filter becomes unable to correctly associate the correct cluster with the truth in the measurement space for subsequent time steps, yielding an inconsistent posterior estimate, as demonstrated by Figure \ref{fig:11kfpgmfail}.
Because the PGM-I update step fails to update the correct cluster in the state space, the use of more information (i.e. observations) in developing an initial state estimate is required, as is a higher-order kinematic fit. As a result, the KF-PGM approach is not effective when few observations are available during that critical first pass.

\begin{figure}[!thpb]
    \centering
    \begin{subfigure}{0.45\textwidth}
        \centering
        \includegraphics[width=\linewidth]{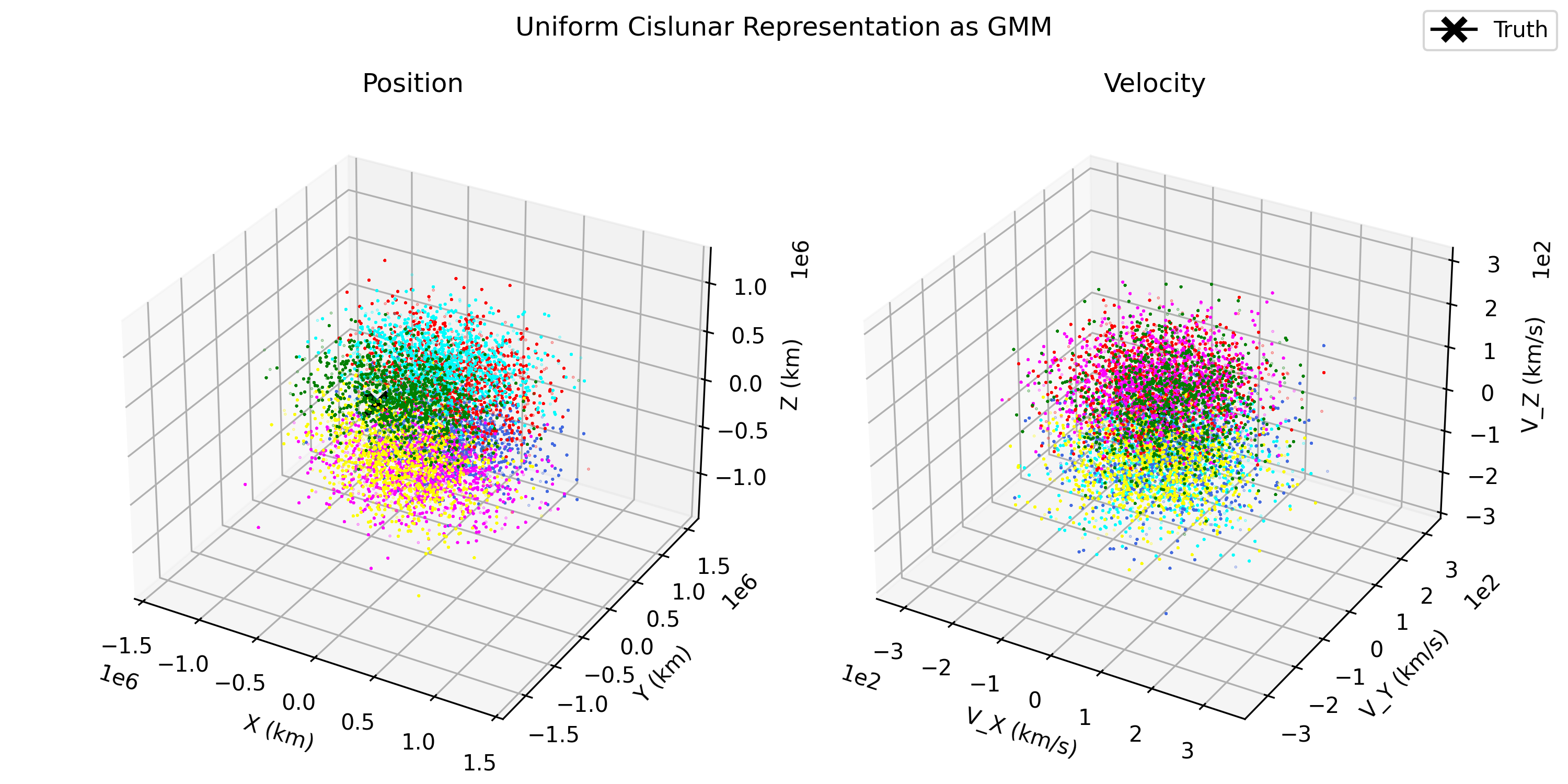}
        \caption{Initial state estimate $p_0^{-}(\mathbf{x})$ that results from clustering the particles in the uniform PDF from Figure \ref{fig:3mvUniform} into a six-component GMM}
        \label{fig:12aPrior}
    \end{subfigure}
    
    \begin{subfigure}{0.45\textwidth}
        \centering
        \includegraphics[width=\linewidth]{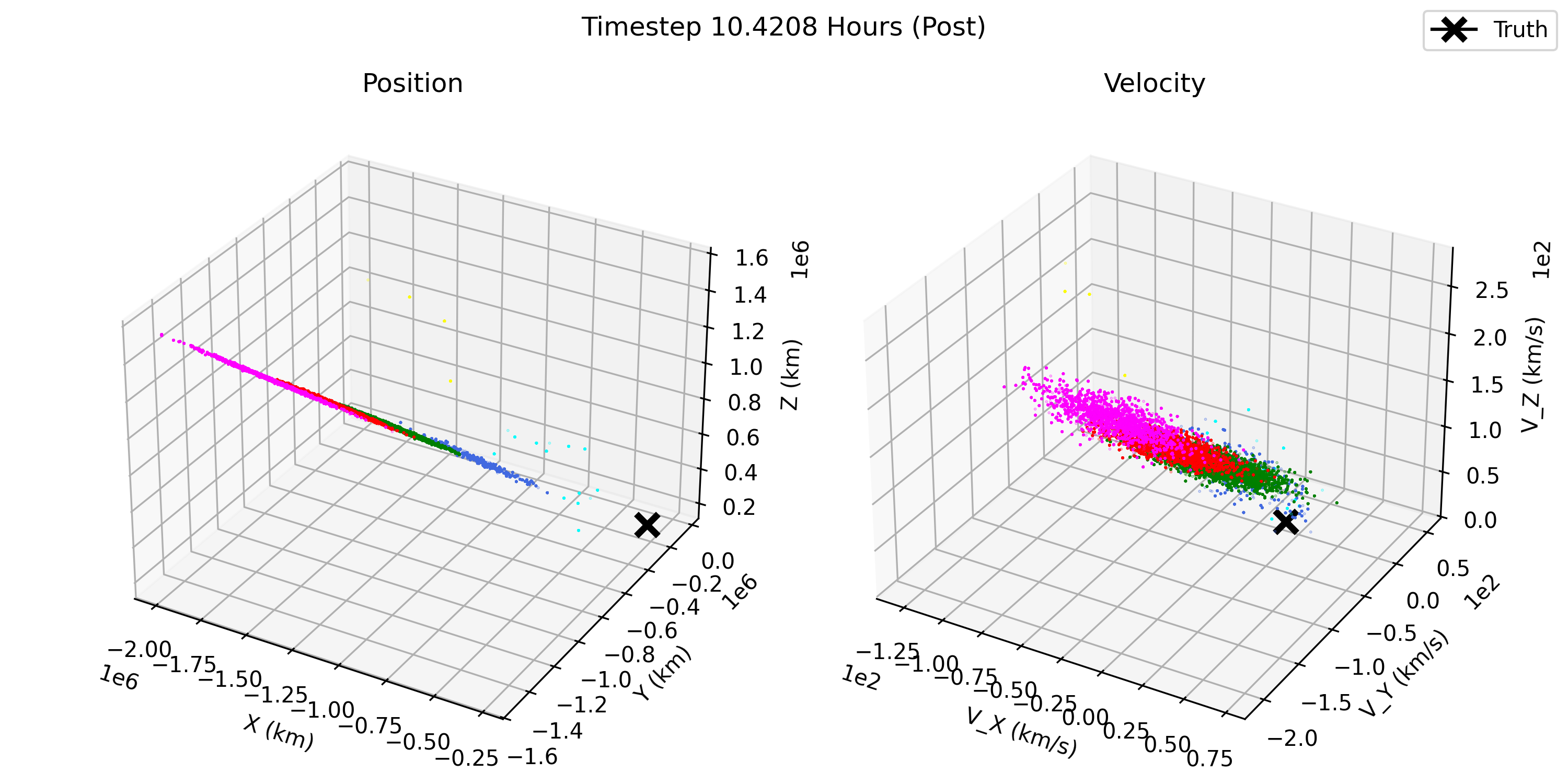}
        \caption{An inconsistent posterior estimate after four iterations of the PGM-I filter}
        \label{fig:12bPosterior}
    \end{subfigure}

    \caption{A demonstration of why a purely recursive PGM-I orbit determination approach is not possible with extremely high initial target uncertainty such as that given by Figure \ref{fig:3mvUniform}. The target estimate becomes inconsistent after just four iterations of the PGM-I filter.}
    \label{fig:12pgm1fail}
\end{figure}

Due to the computationally high cost of MCMC sampling and the PGM-II update step, the question of whether a purely recursive PGM-I orbit determination approach is possible arises. The clustering steps of both PGM filters allow us to model almost any initial or \textit{a priori} estimate as a GMM. Although clustering a uniform PDF will yield a consistent IOD estimate and consistent subsequent filtering estimates, our PGM-I filter will become inconsistent after just four measurement updates, as shown by Figure \ref{fig:12pgm1fail} for a target within the 9:2 resonant NRHO orbit and with an identical measurement set as Section \ref{subsec:ex1}. 
Although Figure \ref{fig:12pgm1fail} partitions the particles within \ref{fig:3mvUniform} into six clusters, higher number of clusters typically result in similar loss of target custody. Because of MCMC sampling, the PGM-II filter is much more efficient than the PGM-I filter at localizing estimates in the measurement space during the first few time steps, when particles in the measurement space are spread across all possible $AZ$ and $EL$ values. KF-PGM's requirement of multiple observations for rendering a robust initial state estimate, coupled with the PGM-I filter's inability to keep custody of targets with large, multivariate uniform initial state estimates underscores the need to utilize a hybrid PGM-based orbit determination approach that makes use of the extended PGM-II filter and PGM-I filter in sequence.

\subsection{Comparison of the H-PGM Solution to Other Hybrid Target Tracking Frameworks}\label{subsec:compsHybrid}

\begin{figure}[h!]
	\centering\includegraphics[width=0.50\columnwidth]{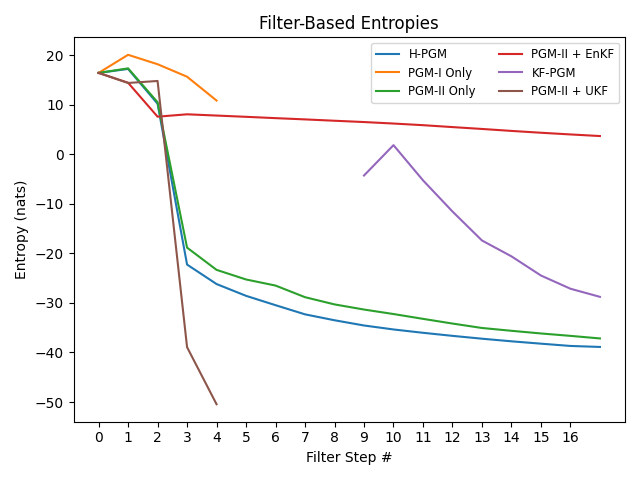}
	\caption{Comparing entropies for several hybrid filters, including a purely recursive PGM-I and PGM-II solutions, and the extended PGM-II filter coupled with a UKF or EnKF.}
	\label{fig:13filterComps}
\end{figure}

The development of a UKF-PF hybrid filter for near-Earth target tracking taught us that the propagation and update steps for any filter may be modularized.\cite{raihanukfpf2018} Similarly, we are not limited to the use of a single filter for orbit determination purposes.\cite{paranjape2025} Just as we switch from the extended PGM-II filter to the PGM-I filter in the hybrid approach after two iterations, we can substitute the PGM-I filter with other, more common filtering approaches. We demonstrate how the entropy changes with each iteration for the same target and measurement set as Section \ref{subsec:ex1} with several hybrid filtering scenarios in Figure \ref{fig:13filterComps}.

Note that the sequences for KF-PGM and H-PGM in Figure \ref{fig:13filterComps} reflect the change in target estimate precision outlined by Table \ref{table:stDevs}, and that the entropy sequence for a purely recursive extended PGM-II approach as that described in Sections \ref{subsec:ex2} and \ref{subsec:ex3} is also provided. The \say{PGM-II + EnKF} sequence may be thought of as the H-PGM sequence, but with the PGM-I filter only containing a single GMM component (i.e. Gaussian \textit{a priori} and posterior estimates). Finally, while the \say{PGM-II + UKF} approach is much faster at reducing state uncertainty, the UKF filter becomes overconfident and loses custody of the target by the fifth filter iteration. The entropy sequence from the purely recursive PGM-I approach from Section \ref{subsec:limitations} is also provided for comparison, demonstrating the importance of the PGM-II filter for the first few update iterations. For many of these hybrid and homogeneous filtering approaches, there is a slight upward tick in the entropy after obtaining the initial step (i.e. Filter Step 0 for most sequences and Filter Step 9 for KF-PGM). The reason for this uptick is the inability of a single angles-only observation set to sufficiently localize the velocity, whose spread may actually widen from the clustering step more than the position will become localized. From the second filtering steps in each sequence onward, target estimates become more precise.

\subsection{Target Characteristic and Measurement Fusion}

\begin{figure}[thpb]
      \centering\includegraphics[width=0.50\columnwidth]{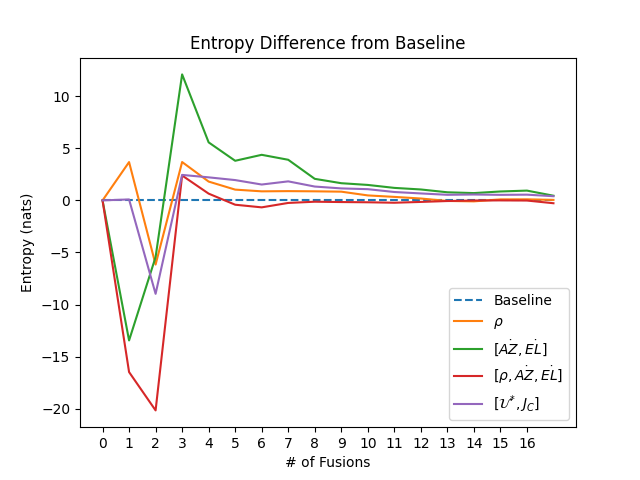}
      \caption{Comparing differences between different information fusion entropy sequences and a baseline entropy sequence over the first pass of a cislunar RSO in the 9:2 NRHO from Section \ref{subsec:ex1}.}
      \label{fig14:secBcomp}
\end{figure}

Throughout this section, we have tracked targets in several cislunar orbit regimes with no \textit{a priori} knowledge of the target's orbit. In reality, we may have more \textit{a priori} information about the cislunar RSO in the form of constraints on characteristics outlined in Table \ref{table:params}. As an example, we can abstract angular rates $[\dot{AZ}, \dot{EL}]^T$ and their constraints into a uniform PDF by recognizing our optical sensor's angular FOV and exposure durations.\cite{bolden2022}. Multiple target characteristics may be vectorized together as \textit{a priori} information, and the resulting multivariate uniform PDFs may be fused in conjunction with the angles-only measurements to further refine the state estimate over time. In this subsection, we fuse all target \textit{a priori} information just once -- after the first $[AZ, EL]^T$ measurement update. Subsequent fusion or filtering steps involve angles-only measurements. At the first time step, choosing whether to fuse angles-only measurements or target characteristic information first will not matter. Subsequent fusions are with angles-only measurements. We compare the initial and subsequent fusions of quantities in Table \ref{table:params} and combinations thereof with the measurements-only (i.e. \say{Baseline}) fusion from Section \ref{subsec:ex1} in Figure \ref{fig14:secBcomp}. Note that the same measurement sequence as Sections \ref{subsec:ex1} and \ref{subsec:compsHybrid} is used. 
    
Each curve in Figure \ref{fig14:secBcomp} represents the entropy difference between the fusions of different sets of target characteristics fused in conjunction with the first measurement relative to the baseline (i.e. measurements-only) fusion. The \say{Baseline} sequence is identical to the \say{H-PGM} sequence in Figure \ref{fig:13filterComps}. The \say{$\rho$} plot refers to range information fusion, namely, knowing that our cislunar RSO is within the 9:2 NRHO at the first time/fusion step. Since target \textit{a priori} information sets such as $[\dot{AZ}, \dot{EL}]^T$ -- corresponding to angular rates fusion -- and $[\rho, \dot{AZ}, \dot{EL}]^T$ -- corresponding to the combination of characteristics fusion -- provide insight into the target velocities, the entropy curves for those fusions do not see an increase during the first 1-2 fusions, unlike those noted Figure \ref{fig:13filterComps}. At the end of the first pass, it becomes clear that having better access to \textit{a priori} information about the cislunar RSO does not result in a materially advantageous precision in the long term. Instead, it shows us that having more \textit{a priori} knowledge is \textit{not} necessary for long-term target tracking, so long as we have a robust filtering framework such as the H-PGM solution. 

\section{Conclusions and Future Work}\label{sec:Conclusions}

In this work, we presented a novel OD framework for cislunar target tracking which involved a small extension of the PGM-II filter to handle initial state estimates and target characteristic information modeled as multivariate uniform PDFs. We emphasized the need to run the PGM-II filter twice for sufficient localization and subsequently switching to the PGM-I filter. Through several examples, we demonstrated the superiority of the H-PGM solution over other hybrid filtering scenarios and purely recursive PGM-I and PGM-II based approaches. We highlighted the PGM-I filter's ability to remove unlikely clusters of \textit{a priori} state estimates after infrequent observations over the PGM-II filter, which tends to retain such clusters for longer periods of time. We also noted that PGM-II filtering is far more computationally intensive than PGM-I filtering due to the MCMC sampling step. Since hybrid PGM-II and UKF or EnKF filtering sequences are either too overconfident or under-confident for simple target tracking scenarios with large uncertainties, the H-PGM solution becomes a viable OD method for cislunar target tracking under extremely high initial uncertainty.

Finally, we observed the short- and long-term effects of fusing cislunar target characteristic information from Table \ref{table:params} with our initial state estimate and the first $[AZ, EL]^T$ measurement set. While fusion of this target \textit{a priori} information yielded some precise estimates in the short term, long term target estimates by the end of our measurement sequence closely matched those of our baseline H-PGM solution. We deduced that while having some \textit{a priori} information about target state might be beneficial, it is not necessarily advantageous in the long term, so long as our filtering framework is robust to minimal information.

We hope to make several extensions to this work. Although we have demonstrated our H-PGM solution for several orbits in the cislunar domain, we wish to test the feasibility of this solution in near-Earth and other xGEO regimes. Additionally, we are eager to extend our H-PGM solution to multi-target tracking scenarios in the cislunar domain due to the effectiveness of the MCMC sampling step in localizing large, uncertain estimates with a single angles-only measurement. Finally, we wish to study the utility of our H-PGM solution with more complex dynamics models such as the small-eccentricity restricted three-body problem (SER3BP) and other, more generalized three- or four-body dynamical systems.

\bibliographystyle{AAS_publication}   
\bibliography{references}   

@book{ussf2020,
    Address = {Washington, DC},
	edition = {1st},
	Publisher = {National Science and Technology Council},
	Title = {National Cislunar Science \& Technology Strategy},
	Year = {2022}}

@article{gooding1997,
    Author = {Gooding, R.H.},
	Journal = {Celest. Mech. Dyn. Astron.},
	Pages = {387--423},
	Title = {A New Procedure for the Solution of the Classical Problem of Minimal Orbit Determination from Three Lines of Sight},
	Volume = {66},
	Number = {4},
	Year = {1997}}

@techreport{taff1979,
     title = {ON GAUSS’S METHOD OF ORBIT DETERMINATION},
     author = {Taff, L.G.},
     group = {94},
     year = {1979},
     institution = {Massachusetts Institute of Technology Lincoln Laboratory},
     month = {06}
}

@article{deMars2013,
author = {DeMars, Kyle J. and Jah, Moriba K.},
title = {Probabilistic Initial Orbit Determination Using Gaussian Mixture Models},
journal = {Journal of Guidance, Control, and Dynamics},
volume = {36},
number = {5},
pages = {1324-1335},
year = {2013},
doi = {10.2514/1.59844},
URL = {https://doi.org/10.2514/1.59844},
eprint = {https://doi.org/10.2514/1.59844}}

@article{kelecy2013,
    Author = {Tom Kelecy and Michael Shoemaker and Moriba Jah},
	Journal = {Proceedings of the 6th European Conference on Space Debris},
	Title = {APPLICATION OF THE CONSTRAINED ADMISSIBLE REGION
            MULTIPLE HYPOTHESIS FILTER TO INITIAL ORBIT
            DETERMINATION OF A BREAK-UP},
	Year = {2013}}

@article{hussein2014,
    Author = {Islam I. Hussein and Christopher W.T. Roscoe and Matthew P. Wilkins and Paul W. Schumacher, Jr.},
	Journal = {Proceedings of the 24th International Symposium on
            Space Flight Dynamics, Laurel, MD},
	Title = {Probabilistic Admissibility in Angles-Only Initial Orbit},
	Year = {2014}}

@article{hussein2014.2,
    Author = {Islam I. Hussein and Christopher W.T. Roscoe and Matthew P. Wilkins and Paul W. Schumacher, Jr.},
	Journal = {Proceedings of the Advanced Maui Optical and Space
            Surveillance Technologies Conference, Wailea, HI},
	Title = {Probabilistic Admissible Region for Short-Arc Angles-Only Observations},
	Year = {2014}}

@article{mishra2024,
  title={Geometric Solution to Probabilistic Admissible Region (PAR)},
  author={Utkarsh Ranjan Mishra and Suman Chakravorty and Weston R. Faber and Islam I. Hussein and Siamak G. Hesar and Benjamin Sunderland},
  journal={The Journal of the Astronautical Sciences},
  year={2024},
  url={https://api.semanticscholar.org/CorpusID:265217760}
}

@article{bolden2022,
    Author = {Mark Bolden and Islam Hussein and Holly Borowski and Robert See and Erin Griggs},
	Journal = {Proceedings of the Advanced Maui Optical and Space
            Surveillance Technologies Conference, Wailea, HI},
	Title = {Probabilistic Initial Orbit Determination and Object Tracking in Cislunar Space Using Optical Sensors},
	Year = {2022}}

@article{griggs2023,
    Author = {Erin Griggs and Matt Schierholtz and Islam Hussein and Mark Bolden and Kyle Charles and Holly Borowski},
	Journal = {Proceedings of the Advanced Maui Optical and Space
            Surveillance Technologies Conference, Wailea, HI},
	Title = {Probabilistic Initial Orbit Determination and Object Tracking in Cislunar Space Using Passive Radio Frequency Sensors},
	Year = {2023}}

@article{raihan2018,
        title = {Particle Gaussian mixture filters-I},
        journal = {Automatica},
        volume = {98},
        pages = {331-340},
        year = {2018},
        issn = {0005-1098},
        doi = {https://doi.org/10.1016/j.automatica.2018.07.023},
        url = {https://www.sciencedirect.com/science/article/pii/S0005109818303807},
        author = {Dilshad Raihan and Suman Chakravorty}}

@book{schaub2003,
	Address = {Reston, VA},
	Author = {Hanspeter Schaub and John L. Junkins},
	Month = {October},
	doi = {10.2514/4.861550},
	Publisher = {{AIAA} Education Series},
	Title = {Analytical Mechanics of Space Systems},
	Year = {2003}}

@mastersthesis{zimovan2017,
    author = {Emily M. Zimovan},
    title = {CHARACTERISTICS AND DESIGN STRATEGIES FOR NEAR RECTILINEAR HALO ORBITS WITHIN THE EARTH-MOON SYSTEM},
    school = {Purdue University},
    year = {2017}
}

@article{zimovanspreen2020,
author={Zimovan-Spreen, Emily M. and Howell, Kathleen C. and Davis, Diane C.},
title={Near rectilinear halo orbits and nearby higher-period dynamical structures: orbital stability and resonance properties},
journal={Celestial Mechanics and Dynamical Astronomy},
year={2020},
month={Jun},
day={13},
volume={132},
number={5},
pages={28},
issn={1572-9478},
doi={10.1007/s10569-020-09968-2},
url={https://doi.org/10.1007/s10569-020-09968-2}
}

@article{mishra2023,
    Author = {Utkarsh R. Mishra and Suman Chakravorty and Islam I. Hussein and Weston Faber and Siamak Hesar and Benjamin Sunderland},
	Journal = {Proceedings of the Advanced Maui Optical and Space
            Surveillance Technologies Conference, Wailea, HI},
	Title = {Comparing Traditional and Admissible-Region Schemes For Angles-Only Initial Orbit},
	Year = {2023}}

@article{koon1999,
    Author = {Wang Sang Koon and Martin W. Lo and Jerrold Marsden and Shane D. Ross},
	Journal = {Proceedings of the International Conference on Differential Equations},
	Title = {Dynamical Systems, the Three-Body Problem, and Space Mission Design},
	Year = {1999}}

@book{holzinger2021,
    Address = {Washington, DC},
	edition = {1st},
	Publisher = {Air Force Research Laboratory},
	Title = {A Primer on Cislunar Space},
	Year = {2021}}

@article{doedel2007,
    author = {DOEDEL, E. J. and ROMANOV, V. A. and PAFFENROTH, R. C. and KELLER, H. B. and DICHMANN, D. J. and GAL\'{A}N-VIOQUE, J. and VANDERBAUWHEDE, A.},
    title = {ELEMENTAL PERIODIC ORBITS ASSOCIATED WITH THE LIBRATION POINTS IN THE CIRCULAR RESTRICTED 3-BODY PROBLEM},
    journal = {International Journal of Bifurcation and Chaos},
    volume = {17},
    number = {08},
    pages = {2625-2677},
    year = {2007},
    doi = {10.1142/S0218127407018671},
    URL = {https://doi.org/10.1142/S0218127407018671},
    eprint = {https://doi.org/10.1142/S0218127407018671}
}

@book{szebehely1969,
    author = {{Szebehely}, Victor and {Grebenikov}, E.},
    title = {Theory of Orbits - The Restricted Problem of Three Bodies},
    publisher = {Academic Press},
    year = 1969
}

@article{raihanukfpf2018,
        title = {An Unscented Kalman-Particle Hybrid Filter for Space Object Tracking},
        journal = {Journal of Astronautical Science},
        volume = {65},
        pages = {111-134},
        year = {2017},
        issn = {},
        doi = {https://doi.org/10.1007/s40295-017-0114-8},
        url = {https://link.springer.com/article/10.1007/s40295-017-0114-8#citeas},
        author = {Dilshad Raihan A.V. and Suman Chakravorty}}

@ARTICLE{lloyd1982,
  author={Lloyd, S.},
  journal={IEEE Transactions on Information Theory}, 
  title={Least squares quantization in PCM}, 
  year={1982},
  volume={28},
  number={2},
  pages={129-137},
  doi={10.1109/TIT.1982.1056489}}

@inproceedings{2019NASA,
  title={National Aeronautics and Space Administration (NASA) White Paper: Gateway Destination Orbit Model: A Continuous 15 Year NRHO Reference Trajectory},
  author={},
  year={2019},
  url={https://api.semanticscholar.org/CorpusID:202637488}
}

@unpublished{paranjape2025,
    author = {Paranjape, Ishan and Chakravorty, Suman},
    title = {CISLUNAR INITIAL ORBIT DETERMINATION AND TARGET TRACKING USING KINEMATIC FITTING AND PARTICLE GAUSSIAN MIXTURE FILTERING},
    note = {to appear in \textit{Proceedings of the 2025 AAS/AIAA Spaceflight Mechanics Conference}. Submitted manuscript accessible at the following \href{https://s3.amazonaws.com/amz.xcdsystem.com/A464D031-C624-C138-7D0E208E29BC4EDD_abstract_File25201/FinalManuscriptPDForDoc_428_0210095240.pdf}{link}.}
}

@unpublished{gupta2025,
    author = {Gupta, Maaninee and DeMars, Kyle},
    title = {CISLUNAR ASTRODYNAMICS LEVERAGING GENERALIZED EQUINOCTIAL ORBITAL ELEMENTS},
    note = {preprint on webpage at \url{https://s3.amazonaws.com/amz.xcdsystem.com/A464D031-C624-C138-7D0E208E29BC4EDD_abstract_File25201/FinalManuscriptPDForDoc_288_0214113128.pdf}}
}

@unpublished{lopez2025,
    author = {Andrea Lopez and Jay MacMahon and Hanspeter Schaub},
    title = {Gaussian Mixture Square Root Filters for Cislunar Angles-Only Relative Orbit Determination},
    note = {preprint on webpage at \url{https://s3.amazonaws.com/amz.xcdsystem.com/A464D031-C624-C138-7D0E208E29BC4EDD_abstract_File25843/PreprintPaperUpload_711_0725041613.pdf}}
}

@article{kalmanBucy1961,
    author = {Kalman, R. E. and Bucy, R. S.},
    title = {New Results in Linear Filtering and Prediction Theory},
    journal = {Journal of Basic Engineering},
    volume = {83},
    number = {1},
    pages = {95-108},
    year = {1961},
    month = {03},
    issn = {0021-9223},
    doi = {10.1115/1.3658902},
    url = {https://doi.org/10.1115/1.3658902},
    eprint = {https://asmedigitalcollection.asme.org/fluidsengineering/article-pdf/83/1/95/5503549/95_1.pdf},
}

@book{crassidis2011,
author = {Crassidis, John L. and Junkins, John L.},
title = {Optimal Estimation of Dynamic Systems, Second Edition (Chapman \& Hall/CRC Applied Mathematics \& Nonlinear Science)},
year = {2011},
isbn = {1439839859},
publisher = {Chapman \& Hall/CRC},
edition = {2nd}
}

@INPROCEEDINGS{wan2000,
  author={Wan, E.A. and Van Der Merwe, R.},
  booktitle={Proceedings of the IEEE 2000 Adaptive Systems for Signal Processing, Communications, and Control Symposium (Cat. No.00EX373)}, 
  title={The unscented Kalman filter for nonlinear estimation}, 
  year={2000},
  volume={},
  number={},
  pages={153-158},
  doi={10.1109/ASSPCC.2000.882463}}

@ARTICLE{julier2004,
  author={Julier, S.J. and Uhlmann, J.K.},
  journal={Proceedings of the IEEE}, 
  title={Unscented filtering and nonlinear estimation}, 
  year={2004},
  volume={92},
  number={3},
  pages={401-422},
  doi={10.1109/JPROC.2003.823141}}

@article{Evensen1994,
  title={Sequential data assimilation with a nonlinear quasi‐geostrophic model using Monte Carlo methods to forecast error statistics},
  author={Geir Evensen},
  journal={Journal of Geophysical Research},
  year={1994},
  volume={99},
  pages={10143-10162},
  url={https://api.semanticscholar.org/CorpusID:16213443}
}

@article{evensen1995,
author = {Evensen, Geir and Van Leeuwen, Peter Jan and A, Edvard},
year = {1995},
month = {06},
pages = {},
title = {Assimilation of Geosat Altimeter Data for the Agulhas Current using the Ensemble Kalman Filter with a Quasi-Geostrophic Model},
volume = {124},
journal = {Monthly Weather Review},
doi = {10.1175/1520-0493(1996)124<0085:AOGADF>2.0.CO;2}
}

@ARTICLE{evensen2003,
       author = {{Evensen}, Geir},
        title = "{The Ensemble Kalman Filter: theoretical formulation and practical implementation}",
      journal = {Ocean Dynamics},
         year = 2003,
        month = nov,
       volume = {53},
       number = {4},
        pages = {343-367},
          doi = {10.1007/s10236-003-0036-9},
       adsurl = {https://ui.adsabs.harvard.edu/abs/2003OcDyn..53..343E}
}

@INPROCEEDINGS{popov2023,
  author={Popov, Andrey A and Zanetti, Renato},
  booktitle={2023 26th International Conference on Information Fusion (FUSION)}, 
  title={Ensemble Gaussian Mixture Filtering with Particle-localized Covariances}, 
  year={2023},
  volume={},
  number={},
  pages={1-7},
  doi={10.23919/FUSION52260.2023.10224227}}

@article{Iannamorelli2025AdaptiveGM,
  title={Adaptive Gaussian Mixture Filtering for Multi-sensor Maneuvering Cislunar Space Object Tracking},
  author={John L. Iannamorelli and Keith A. LeGrand},
  journal={The Journal of the Astronautical Sciences},
  year={2025},
  url={https://api.semanticscholar.org/CorpusID:275418755}
}

@article{heidrich2025,
    title={Universal Angles-Only Cislunar Initial Orbit Determination Using Sparse Grid Collocation},
    author={Casey Heidrich and Marcus Holzinger},
    journal={The Journal of the Astronautical Sciences},
    year={2025},
    url={https://link.springer.com/article/10.1007/s40295-025-00491-w}
}

@misc{paranjape2026,
      title={Probabilistic Methods for Initial Orbit Determination and Orbit Determination in Cislunar Space}, 
      author={Ishan Paranjape and Tarun Hejmadi and Suman Chakravorty},
      year={2026},
      eprint={2602.18058},
      archivePrefix={arXiv},
      primaryClass={astro-ph.EP},
      url={https://arxiv.org/abs/2602.18058}, 
}

@inproceedings{tapley2004,
  title={Statistical Orbit Determination},
  author={Byron D. Tapley and Bob E. Schutz and George H. Born},
  year={2004},
  url={https://api.semanticscholar.org/CorpusID:267046698}
}

@article{yun2022,
    title = {Kernel-based ensemble gaussian mixture filtering for orbit determination with sparse data},
    journal = {Advances in Space Research},
    volume = {69},
    number = {12},
    pages = {4179-4197},
    year = {2022},
    issn = {0273-1177},
    doi = {https://doi.org/10.1016/j.asr.2022.03.041},
    author = {Sehyun Yun and Renato Zanetti and Brandon A. Jones}
}

@article{durant2023,
    Author = {Dalton Durant and Andrey Popov and Renato Zanetti},
	Journal = {Proceedings of the AAS/AIAA Astrodynamics Specialist                Conference, Austin, TX},
    Title = {MCMC EnGMF for Sparse Data Orbit Determination},
	Year = {2023}}

@article{raihan2018pgm2,
    title = {Particle Gaussian mixture filters-II},
    journal = {Automatica},
    volume = {98},
    pages = {341-349},
    year = {2018},
    issn = {0005-1098},
    doi = {https://doi.org/10.1016/j.automatica.2018.07.024},
    author = {Dilshad Raihan and Suman Chakravorty}
}

@article{givens2025,
    Author = {Matthew Givens and Alexandre Cortiella and Amanda Marlow and Veronica Rankowicz and Andrew Koehler and Aaron Liao and Justin Spurbeck and Charles Cain and Patrick Miga},
	Journal = {Proceedings of the Advanced Maui Optical and Space
            Surveillance Technologies Conference, Wailea, HI},
	Title = {Novel Algorithms for Custody Maintenance and Tracking in the Cislunar Environment},
	Year = {2025}}

@INPROCEEDINGS{raihan2018fusion,
  author={Raihan, D. and Chakravorty, S.},
  booktitle={2018 21st International Conference on Information Fusion (FUSION)}, 
  title={Particle Gaussian Mixture Filters-II}, 
  year={2018},
  volume={},
  number={},
  pages={1092-1099},
  doi={10.23919/ICIF.2018.8455307}}

@article{Peterson2023,
  title={Local orbital elements for the circular restricted three-body problem},
  author={Peterson, Luke T and Scheeres, Daniel J},
  journal={Journal of Guidance, Control, and Dynamics},
  volume={46},
  number={12},
  pages={2275--2289},
  year={2023},
  publisher={American Institute of Aeronautics and Astronautics}
}

@mastersthesis{dahlke2018,
    author = {Jacob A. Dahlke},
    title = {OPTIMAL TRAJECTORY GENERATION IN A DYNAMIC MULTI-BODY
            ENVIRONMENT USING A PSEUDOSPECTRAL METHOD},
    school = {Air Force Institute of Technology},
    year = {2018}
}

@article{kim2026,
    title = {DBSCAN-based particle Gaussian mixture filters},
    journal = {Digital Signal Processing},
    volume = {168},
    pages = {105546},
    year = {2026},
    issn = {1051-2004},
    doi = {https://doi.org/10.1016/j.dsp.2025.105546},
    url = {https://www.sciencedirect.com/science/article/pii/S1051200425005688},
    author = {Sukkeun Kim and Mengwei Sun and Ivan Petrunin and Hyo-Sang Shin}
}

@inproceedings{spreen2017,
    author = {Spreen, Emily and Howell, Kathleen and Davis, Diane},
    year = {2017},
    month = {05},
    pages = {},
    title = {Near Rectilinear Halo Orbits and Their Application in Cis-lunar Space}
}

@article{mishra2021,
    Author = {Utkarsh Ranjan Mishra and Weston Faber and Suman Chakravorty and Islam Hussein},
	Journal = {Proceedings of the Advanced Maui Optical and Space
            Surveillance Technologies Conference, Wailea, HI},
	Title = {A Subset Simulation based technique for calculating the probability of collision},
	Year = {2021}}

\end{document}